\documentclass[pdftex,twocolumn,epjc3]{svjour3}
\RequirePackage[T1]{fontenc}
\smartqed  

\RequirePackage{graphicx}
\RequirePackage{mathptmx}      
\RequirePackage{flushend}
\RequirePackage[numbers,sort&compress]{natbib}
\RequirePackage[colorlinks,citecolor=blue,urlcolor=blue,linkcolor=blue]{hyperref}

\usepackage[MeX,plmath]{polski}
\usepackage[utf8]{inputenc}
\usepackage{float}
\usepackage[polish,USenglish]{babel}
\usepackage{enumitem}
\usepackage{sidecap}
\usepackage{caption}
\usepackage{subcaption}
\usepackage{amsmath}
\usepackage{braket}
\usepackage{multirow}

\usepackage{array}
\newcolumntype{L}[1]{>{\raggedright\let\newline\\\arraybackslash\hspace{0pt}}m{#1}}
\newcolumntype{C}[1]{>{\centering\let\newline\\\arraybackslash\hspace{0pt}}m{#1}}
\newcolumntype{R}[1]{>{\raggedleft\let\newline\\\arraybackslash\hspace{0pt}}m{#1}}

\usepackage{pgfplots}
\usetikzlibrary{spy}

\newcommand{\ops}{$\mbox{o-Ps}\rightarrow 3\gamma$ }
\newcommand{\sigmaThetaPointLike}{0.4^{\circ}}
\newcommand{\sigmaThetaGps}{4.2^{\circ}}
\newcommand{\sigmaEnergyPointLike}{4.1 \mbox{ keV}}
\newcommand{\sigmaEnergyGps}{30 \mbox{ keV}}

\journalname{Eur. Phys. J. C}

\begin{document}

\title{A feasibility study of ortho-positronium decays measurement with the J-PET scanner based on plastic scintillators}

\author{
    D.~Kamińska\thanksref{WFAIS}
    \and
    A.~Gajos\thanksref{WFAIS} 
    \and
    E.~Czerwiński\thanksref{WFAIS} 
    \and
    D.~Alfs\thanksref{WFAIS} 
    \and
    T.~Bednarski\thanksref{WFAIS} 
    \and
    P.~Białas\thanksref{WFAIS}
    \and
    C.~Curceanu\thanksref{LNF}
    \and
    K.~Dulski\thanksref{WFAIS} 
    \and
    B.~Głowacz\thanksref{WFAIS} 
    \and
    N.~Gupta-Sharma\thanksref{WFAIS} 
    \and
    M.~Gorgol\thanksref{UMCS} 
    \and
    B.~C.~Hiesmayr\thanksref{Vienna}
    \and
    B.~Jasińska\thanksref{UMCS}
    \and
    G.~Korcyl\thanksref{WFAIS}
    \and
    P.~Kowalski\thanksref{SWIERK} 
    \and
    W.~Krzemień\thanksref{SWIERKHEP}
    \and
    N.~Krawczyk\thanksref{WFAIS}
    \and
    E.~Kubicz\thanksref{WFAIS}
    \and
    M.~Mohammed\thanksref{WFAIS}
    \and
    Sz.~Niedźwiecki\thanksref{WFAIS} 
    \and
    M.~Pawlik-Niedźwiecka\thanksref{WFAIS} 
    \and
    L.~Raczyński\thanksref{SWIERK} 
    \and
    Z.~Rudy\thanksref{WFAIS} 
    \and
    M.~Silarski\thanksref{LNF} 
    \and
    A.~Wieczorek\thanksref{WFAIS} 
    \and
    W.~Wiślicki\thanksref{SWIERK}
    \and
    B.~Zgardzińska\thanksref{UMCS} 
    \and
    M.~Zieliński\thanksref{WFAIS} 
    \and
    P.~Moskal\thanksref{WFAIS}
}
\institute{Faculty of Physics, Astronomy and Applied Computer Science, Jagiellonian University,  S.~Łojasiewicza 11, 30-348 Kraków, Poland\label{WFAIS}
    \and
    INFN, Laboratori Nazionali di Frascati CP 13,  Via E. Fermi 40, 00044, Frascati, Italy\label{LNF}
    \and
    Faculty of Physics, University of Vienna  Boltzmanngasse 5, 1090 Vienna, Austria\label{Vienna}
    \and
    Department of Nuclear Methods, Institute of Physics, Maria Curie-Sklodowska University, Pl.~M.~Curie-Sklodowskiej~1, 20-031 Lublin, Poland\label{UMCS}
    \and
    Świerk Computing Centre, National Centre for Nuclear Research,  05-400 Otwock-Świerk, Poland\label{SWIERK}
    \and
    High Energy Department, National Centre for Nuclear Research,  05-400 Otwock-Świerk, Poland\label{SWIERKHEP}
}            
\maketitle
\begin{abstract}
We present a  study of the application
of the Jagiellonian Positron Emission Tomograph (J-PET) 
for the registration of gamma quanta from decays of ortho-positronium (o-Ps).
The J-PET is the first positron emission tomography scanner based on organic scintillators 
in contrast to all current PET scanners based on inorganic crystals. 
Monte Carlo simulations show that the J-PET as an axially symmetric and high acceptance scanner 
can be used as a multi-purpose detector
well suited to pursue research including e.g. tests of discrete symmetries in decays of ortho-positronium
in addition to the medical imaging.
     \\
The gamma quanta originating from o-Ps decay interact in the plastic
scintillators predominantly via the Compton effect, 
making the direct measurement of their energy impossible.
Nevertheless, it is shown in this paper that the J-PET scanner will 
enable studies of the \ops decays
with angular and energy resolution
equal to $\sigma(\theta) \approx \sigmaThetaPointLike$ and $\sigma(E) \approx \sigmaEnergyPointLike$,
respectively.
An order of magnitude shorter decay time of signals from plastic scintillators
with respect to the inorganic crystals
results  not only in better timing properties  crucial for the reduction of physical and instrumental background,
but also suppresses significantly the pile-ups,
thus enabling compensation of the lower efficiency of 
the plastic scintillators by performing measurements with higher positron source activities. 
\end{abstract}
\section{Introduction}
\label{sec:introduction}
The Positron Emission Tomography (PET) is based on registration of two gamma quanta originating from a 
positron annihilation in matter.
However, the $e^+ e^- \rightarrow 2 \gamma$  process is not the only possible route of positron annihilation. 
Electron and positron may annihilate also to a larger number of gamma quanta
with lower probability, or form a bound state called positronium.
In the ground state with angular momentum equal to zero positronium may be formed in the triplet state (with spin S~=~1) 
referred to as ortho-positronium (o-Ps), or singlet state (S~=~0) 
referred to as para-positronium (p-Ps). Positronium, being a bound-state built from electron and anti-electron bound 
by the central potential, is an eigenstate of both charge (C) 
and spatial parity~(P) operators, as well as of their combination (CP). 
Therefore, it is  well suited  for the studies of these discrete 
symmetries in the leptonic sector. These symmetries may be studied by the measurement of the 
expectation values of various operators (odd with respect to the studied symmetry)
constructed from the momenta of photons and the spin of the ortho-positronium~\cite{Moskal:2016moj}. 
Such studies are limited by the photon-photon interaction,  
however it was estimated that the vacuum polarisation effects may mimic the CP and CPT symmetries 
violation only at the level of 10$^{-9}$~\cite{Bernreuther:1988tt}, 
which is still by six orders of magnitude less than the presently best known experimental 
limits for CP and CPT violations in the positronium decays which are at the level of 0.3\%~\cite{gammasphere,tokyo}.
Ortho-positronium is symmetric in space and spin and, therefore, as a system built from fermions
it must be charge symmetry odd. Para-positronium, in turn, as anti-symmetric in spin and symmetric in space, must be charge symmetry even. C symmetry conservation implies that the 
ortho-positronium annihilate into 
odd number of gamma quanta, $3\gamma$ being the most probable, with lifetime 142~ns and para-positronium decays into 
even number of gamma quanta with  lifetime 125~ps~\cite{Harpen:2003zz,Ramadhan,Vallery,Jinnouchi}. Such a huge difference in the life-times
enables an efficient experimental disentangling of  o-Ps from p-Ps decays. 

\begin{figure}[h!]
    \centering
    \includegraphics[width=0.5\textwidth]{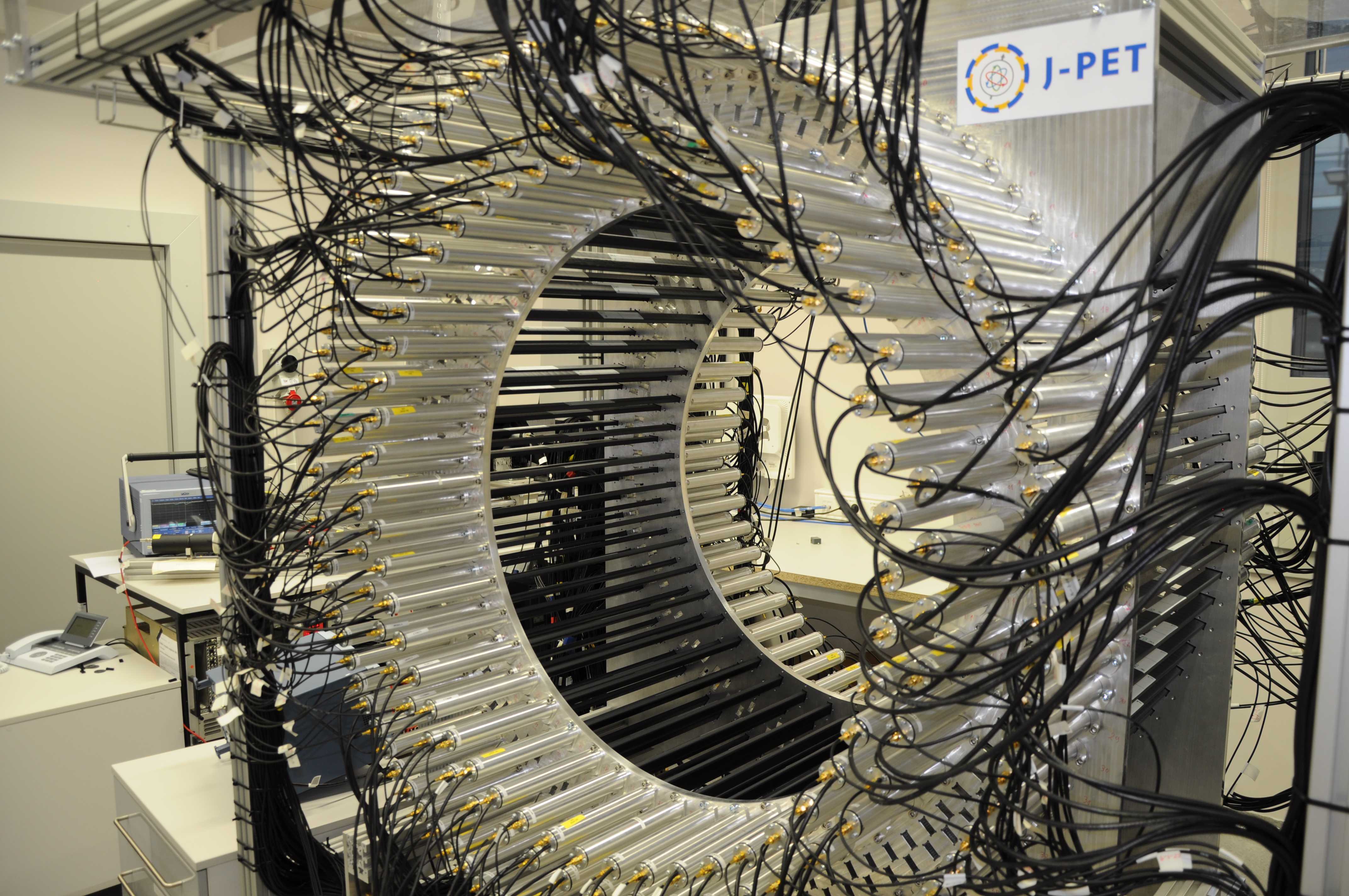}
    \caption{Photo of the Jagiellonian Positron Emission  Tomograph  (J-PET). The J-PET detector is made of three cylindrical layers of
        EJ-230 plastic scintillator strips (black) with dimension of $7\times 19 \times 500 \mbox{ mm}^3$  
        and Hamamatsu R9800  vacuum tube photomultipliers (grey). The signals from photomultipliers are probed in the
        voltage domain at four thresholds with the timing accuracy of about 30~ps~\cite{Palka:2014} and the data acquisition is working in the
    trigger-less mode~\cite{Korcyl:2014,Korcyl:2016pmt}. }
    \label{fig:foto_det}
\end{figure}

\begin{figure}[h!]
    \centering
        \includegraphics[width=\columnwidth]{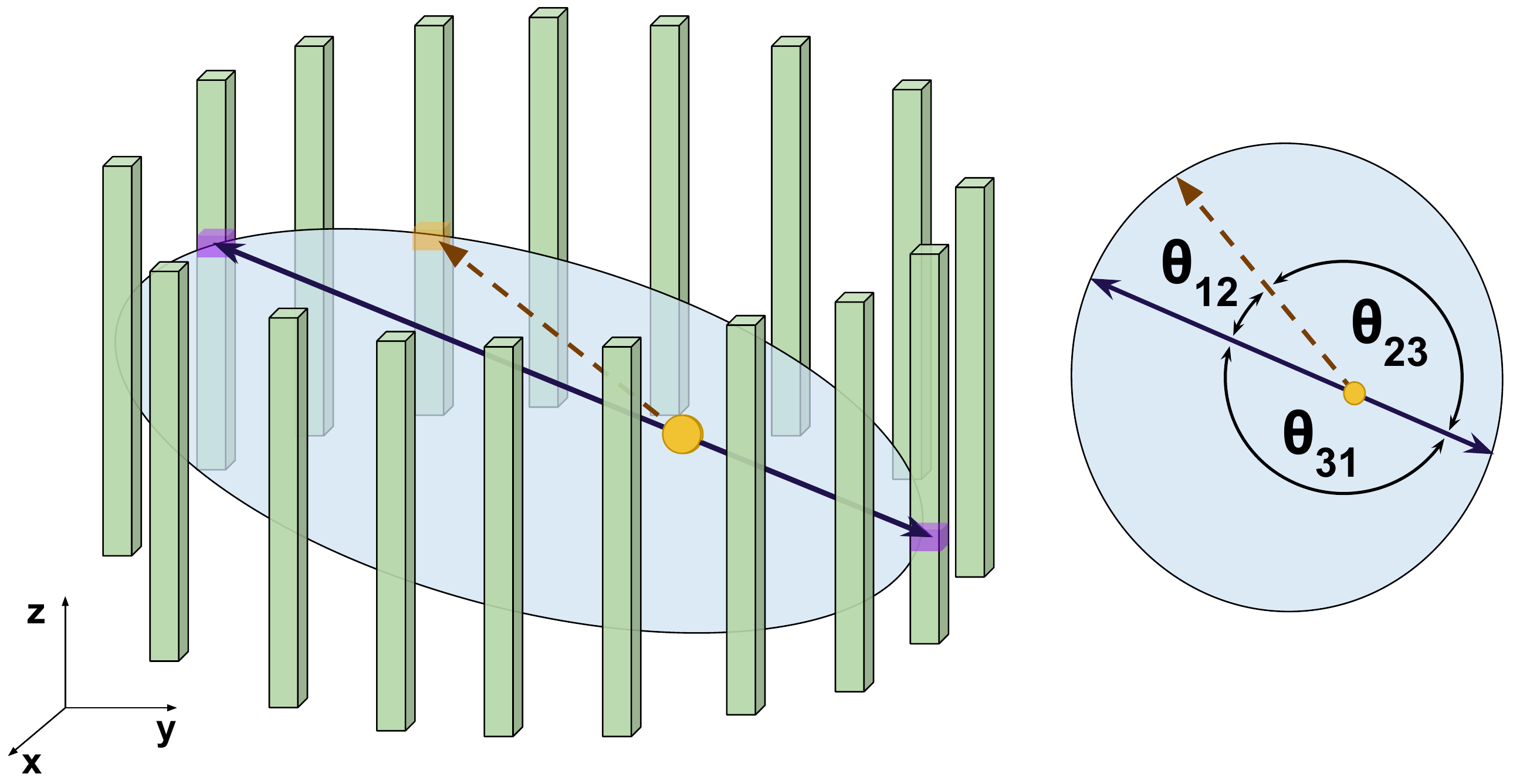}
        \includegraphics[width=\columnwidth]{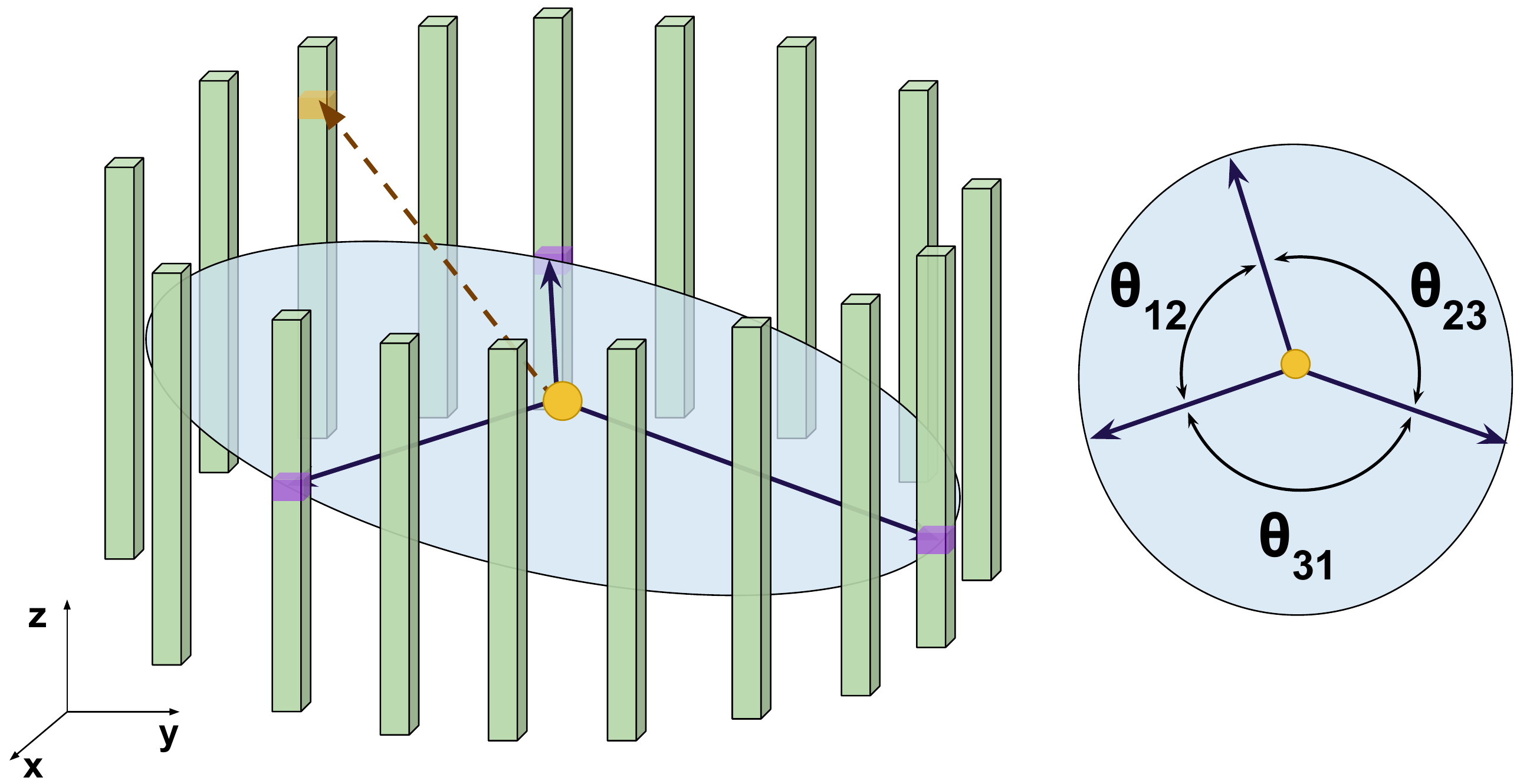}
    \caption{Schematic view of a single layer of the J-PET detector with 
        two (up) or three (down) gamma quanta annihilation.
        In presently built geometry the first layer consists of 48 plastic scintillators (green bars).
        In this pictorial representation, for  clarity, a smaller number of strips is shown.
        Solid dark blue lines indicate annihilation quanta and dashed brown line indicates 
        de-excitation gamma quantum e.g. from the 
        $^{22} \mbox{Na} \rightarrow  \ ^{22}\mbox{Ne}^* + e^+ + \nu \rightarrow \ ^{22}\mbox{Ne} +  \gamma + e^+ + \nu$ 
        decay chain.
        Due to the momentum conservation annihilation quanta are moving along the same line 
        in the case of $e^+e^- \rightarrow 2\gamma$, while  in the case of the $e^+e^- \rightarrow 3\gamma$ 
        they are included in a single plane.  The de-excitation photon (dashed line) is not correlated 
        with the annihilation photons and is isotropically distributed with respect to the
        annihilation Plane-Of-Response.
        Due to the fact that annihilation and de-excitation occur in a good approximation at the same place 
        the photons from the $e^+e^- \rightarrow 2\gamma$ form a plane with the de-excitation photon.  
    }   
    \label{scheme_detector}
\end{figure}

With the recently constructed J-PET detector (see Figure~\ref{fig:foto_det}) we intend to study the \ops process in order to 
examine discrete symmetries and to test new medical imaging techniques based on the detection of three photons~\cite{patent}.
In the ortho-positronium decay the additional information carried by the  $3^{rd}$ $\gamma$  allows  more precise annihilation point reconstruction. 
Schematic view of p-Ps and o-Ps annihilation is shown in Figure~\ref{scheme_detector}.
Moreover, the observed yield of three
gamma annihilation depends on material's properties (see Subsection~\ref{sec:formation}),
therefore it may allow to gain some information not only about location but also about  properties of tumors~\cite{Kacperski}.
In fundamental physics, studies of the three gamma annihilation allows 
not only to test the discrete symmetry 
violation~\cite{Kaminska:2015yqa} 
but also enables searches of physics beyond the Standard Model: extra dimensions~\cite{Dubovsky}, dark matter~\cite{Crivelli}
and a new light vector gauge boson~\cite{Gninenko}.
Since a detailed physics program of J-PET and its motivation is described elsewhere in a dedicated article~\cite{Moskal:2016moj},
here as an example we would like only to discuss briefly experimental approach to determining the expectation 
value of the odd operator for the CPT symmetry, whose violation has not been observed so far. As it was recently
shown~\cite{Gajos:2016nfg} the \mbox{J-PET} detector allows for a spin direction ($\vec{S}$) determination of o-Ps 
created in cylindrical target. 
Additionally as it is described in Section~\ref{sec:results} the J-PET detector enables determination of 
the momentum vectors of gamma quanta originating from the \ops process.
These properties allow for construction of
the following operator odd under CPT transformation: $\vec{S}\cdot(\vec{k_1}\times\vec{k_2})$,
where $\vec{k_1}$ and $\vec{k_2}$
denote momenta of the most and second most energetic quanta, respectively. 
The non-zero expectation value (indicating violation of CPT symmetry) would 
manifest itself as an asymmetry between numbers of events with spin direction pointing to opposite sides of the decay plane ($\vec{k_1}\times\vec{k_2}$).
In this paper we focus on the feasibility study of the detection of o-Ps annihilation.

The registration of three-gamma annihilation and conducting  of the above mentioned  research
is possible by the J-PET detector whose  novelty lies in application of plastic scintillators instead of crystals~\cite{patent_jpet}.
This solution allows to sample fast signals ($5$ ns)~\cite{Moskal:2016ztv,Moskal:2014rja,Raczynski:2015zca,Raczynski:2014poa,Palka:2014}
and build more extended geometries, in comparison
to commercially used PET detectors~\cite{Moskal:2016ztv}.
In this paper we study the feasibility of the three gamma annihilation 
measurements using the J-PET detector. To this end we have developed 
Monte Carlo simulations accounting~for:
\begin{enumerate}[label=(\roman*)]
    \item positron emission and thermalisation in the target material,
    \item angular and energy distributions  of gamma quanta originating from 
        ortho-positronium annihilation,
    \item Compton interactions of emitted gamma quanta in the detector built from plastic scintillators, 
    \item determination of gamma quanta hit-position and hit-time in the detector with 
        experimentally determined resolutions,
    \item multiple scattering and accidental coincidences, 
    \item reconstruction of  registered gamma quanta four-momenta, 
\end{enumerate}
and used four possible geometrical configurations  of the J-PET detector.

Section \ref{sec:MonteCarlo} gives a general introduction of positron emission 
and interaction with matter together with the formation of positronium 
and the description of ortho-positronium annihilation into three gamma quanta. 
Possible detector geometries are summarized in Section~\ref{sec:MC_det}.
Properties of J-PET detector, comparison between simulated and experimental spectra 
and the method of background rejection are presented in Section~\ref{sec:det}.
Section \ref{sec:results} contains the detector
efficiency estimation as well as the energy and angular resolutions. 
\section{Performance assessment: Monte Carlo simulations}
\label{sec:MonteCarlo}
The following paragraphs contain the description of Monte Carlo simulations of positrons
emitted from $\beta^+$ source ($^{22}$Na)
that bind with electron and form
positronium. Simulation takes into account the effects of finite positronium 
range and non-zero residual momentum of the
annihilation positron-electron pair. 
Special emphasis is put on a  proper description of available phase-space of photons from 
ortho-positronium annihilation and their further detection in the J-PET detector that consists of plastic
scintillators.
\subsection{Positron source and positronium formation}
\label{sec:formation}
\begin{figure}[b]
    \centering
    \includegraphics[width=0.8\columnwidth]{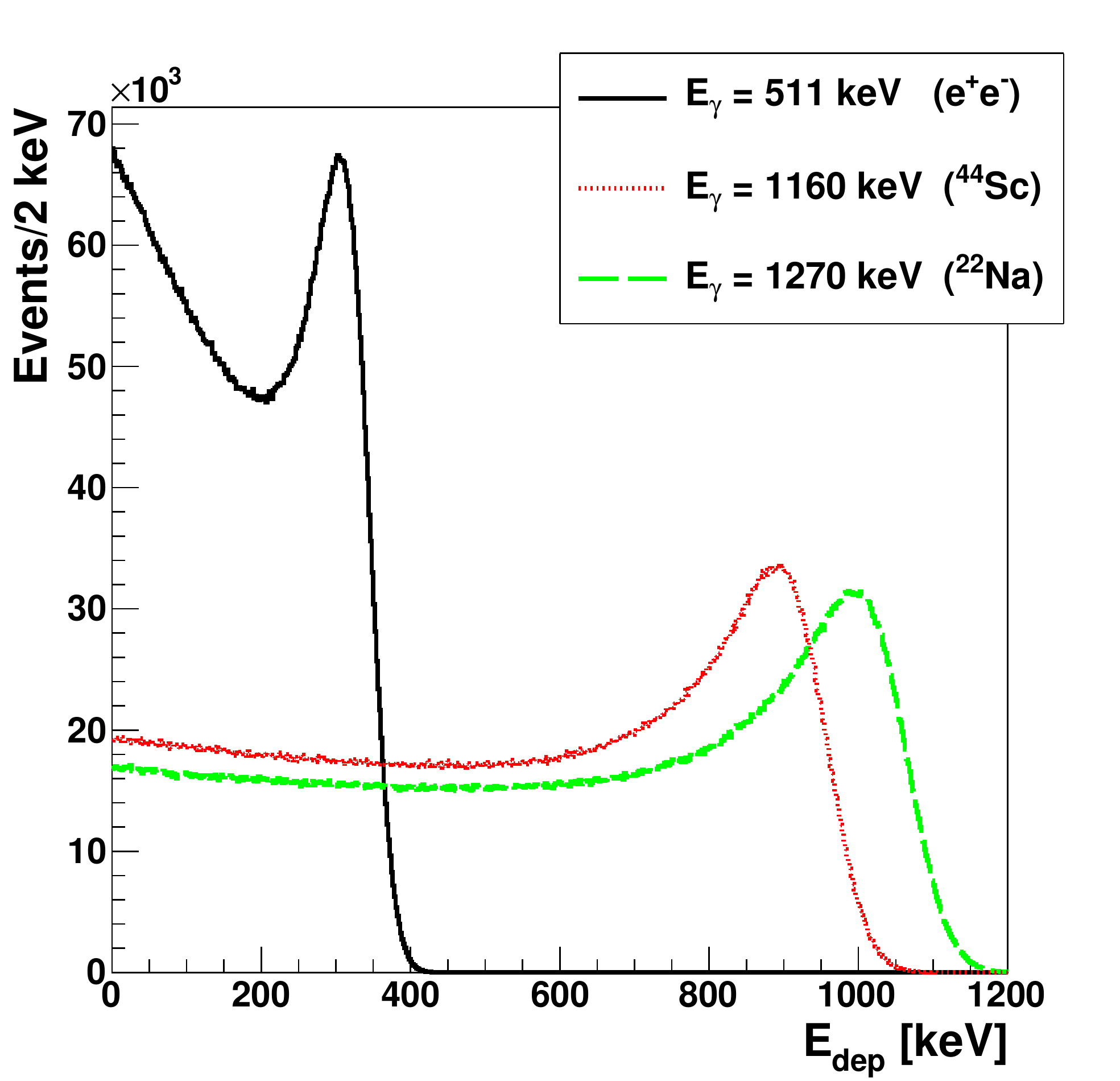}
    \caption{Simulated spectra of deposited energy in plastic scintillators for gamma quanta 
        from $e^+ e^- \rightarrow 2 \gamma$ annihilation and  for de-excitation 
        gamma quanta originating from isotopes indicated in the legend.
        The spectra were simulated including the energy resolution of the J-PET detector \cite{Moskal:2016ztv}
    and were normalized to the same number of events.}
    \label{Prompt_gamma_compton}
\end{figure}

Table~\ref{table_isotopes} summarizes the important characteristics of the isotopes used for 
different types of imaging as well as in laboratory studies. 
Those isotopes decay through $\beta^+$ transitions emitting a positron  that travels through matter, scatters 
and slows down reaching thermal energies. Then it
undergoes free annihilation or forms a positronium~\cite{positron_chemistry}.
In water at $20^{\circ}$C the positron has about $64\%$ chance of undergoing free annihilation~\cite{Colombino}. 
The positronium is produced mostly in the ground state
forming para-positronium ($^1S_0$, p-Ps) or ortho-positronium ($^3S_0$, o-Ps)
with probability of $25\%$ and $75\%$, respectively.
The annihilation of those states is leading predominantly to an emission of two or three gamma quanta for
p-Ps or o-Ps states, respectively. However, the interactions with matter  can lead to inversion 
of the ortho-positronium spin or to the pick-off processes and, as a result,
can affect the relative ratio of $3\gamma/2\gamma$ annihilation.
The effective yield of annihilation into $3\gamma$  in most of non-metallic substances  is of the order
of~$1\%$, although in some cases, as for example fine powders of alkaline oxides, it can reach 
even 29\% as recently shown for the  amberlite porous polymer XAD-4 (CAS 37380-42-0)~\cite{Jasinska:2016qsf}. 
\begin{table*}
    \centering
    \caption{Summary of major physical characteristics of beta-plus isotopes useful for PET imaging and  
        positron annihilation lifetime spectroscopy (PALS)
        investigations.
        For isotopes that decay into 
        excited states the properties of emitted gamma quanta are denoted.
    Data were adapted from~\cite{nncd}.}
    \label{table_isotopes}
        \begin{tabular*}{\textwidth}{@{\extracolsep{\fill}}|c|c|c|c|c|c|@{}} \hline
            Isotope & Half-life & $\beta^+$ decay & $E_{\gamma}$ [MeV] &
            $E_{e^+}^{max}$ [MeV] & Excited nuclei lifetime \\ \hline
            \multicolumn{6}{|c|}{Isotopes for PALS and PET imaging} \\\hline
            $^{22}$Na & 2.6 [year] & $^{22} \mbox{Na} \to ^{22}\mbox{Ne} + e^+ + \nu_e +\gamma$
            & 1.27 & 0.546 &  3.63~[ps] \\ \hline
            $^{68}$Ga & 67.8 [min] &  $^{68}\mbox{Ga} \to ^{68}\mbox{Zn} + e^+ + \nu_e +\gamma$
            & 1.08& 0.822 & 1.57 [ps] \\ 
            $^{44}$Sc & 4.0 [h] & $^{44}\mbox{Sc} \to ^{44}\mbox{Ca} + e^+ + \nu_e +\gamma$ &
            1.16 & 1.474 & 2.61 [ps] \\ \hline 
            \multicolumn{6}{|c|}{ Isotopes for PET imaging} \\ \hline  
            $^{68}$Ga & 67.8 [min] &  $^{68}\mbox{Ga} \to ^{68}\mbox{Zn} + e^+ + \nu_e$
            & - & 1.899 & - \\ 
            $^{11}$C & 20.4 [min] & $^{11}\mbox{C} \to ^{11}\mbox{B}+ e^+ + \nu_e $
            & - &  0.961 & - \\
            $^{13}$N & 10.0 [min] & $^{13}\mbox{N} \to  ^{13}\mbox{C}+ e^+ + \nu_e$
            & - & 1.198 & - \\
            $^{15}$O & 2.0 [min] & $^{15}\mbox{O} \to  ^{15}\mbox{N}+ e^+ + \nu_e$ 
            & - & 1.735 & - \\
            $^{18}$F &  1.8 [h] & $^{18}\mbox{F} \to  ^{18}\mbox{O} + e^+ + \nu_e$
            & - & 0.634 & - \\ \hline
        \end{tabular*}
\end{table*}
Some of the $\beta^+$ emitters, e.g. $^{22}$Na or $^{44}$Sc, decay to daughter nucleus in 
excited states and emit prompt gamma with a well defined energy.
In plastic scintillators gamma quanta interact mostly via the Compton
scattering.
Figure~\ref{Prompt_gamma_compton} shows the energy loss spectrum expected for the gamma quanta 
from the $e^+e^- \rightarrow 2 \gamma$ annihilation compared to the spectra expected from the de-excitation 
quanta from $^{22}$Na and $^{44}$Sc isotopes.
The results were obtained taking into account the experimental energy resolution of the J-PET detector~\cite{Moskal:2014sra}. 
The identification of de-excitation and annihilation photons is based on the energy loss and angular correlations. 
Using the energy loss criterion (e.g. $E_{dep} > 0.370$~MeV) we can uniquely identify de-excitation quantum from 
the $^{44}$Sc and $^{22}$Na decays with a selection efficiency of 0.66 and 0.70, respectively.
The second selection method is, however,  much more efficient. It will be based on the relation between the relative angles of the photons directions.
The trilateration  method allows reconstruction of an emission point~\cite{Gajos:2016nfg} and 
the relative angles between the gamma quanta.
After assigning the numbers to the photons such that the relative angles are arranged in the ascending order ($\theta_{12} < \theta_{23} < \theta_{31}$),
in the case of the 2$\gamma$ annihilation (Figure~\ref{scheme_detector}  left) the largest angle $\theta_{31}$  will be equal to 180 degrees and will correspond to the photons from the
$e^+ e^- \rightarrow 2 \gamma$ process.
Therefore, the de-excitation gamma quantum can be identified as photon number 2. 
This second selection method is independent of the energy loss criteria, and due to the high angular resolution
of the J-PET tomograph (see Section \ref{sec:MC_det}), 
it will allow an identification with close to 100\% selection efficiency.  
In case of the 3$\gamma$ annihilation, photons originating from \ops 
process are emitted in a single plane. The gammas directions are not 
correlated with the de-excitation photon (see Figure~\ref{scheme_detector}),
so probability of their miss-identification with the 
de-excitation gamma quanta  
is at the level of few percent only, 
and for long lifetime of o-Ps (larger than few ns) it is negligible due to the large difference
between the hit-times of annihilation and de-excitation photons which may be used as additional third criterion.

\begin{figure*}[p]
    \centering
    \includegraphics[width=.9\textwidth]{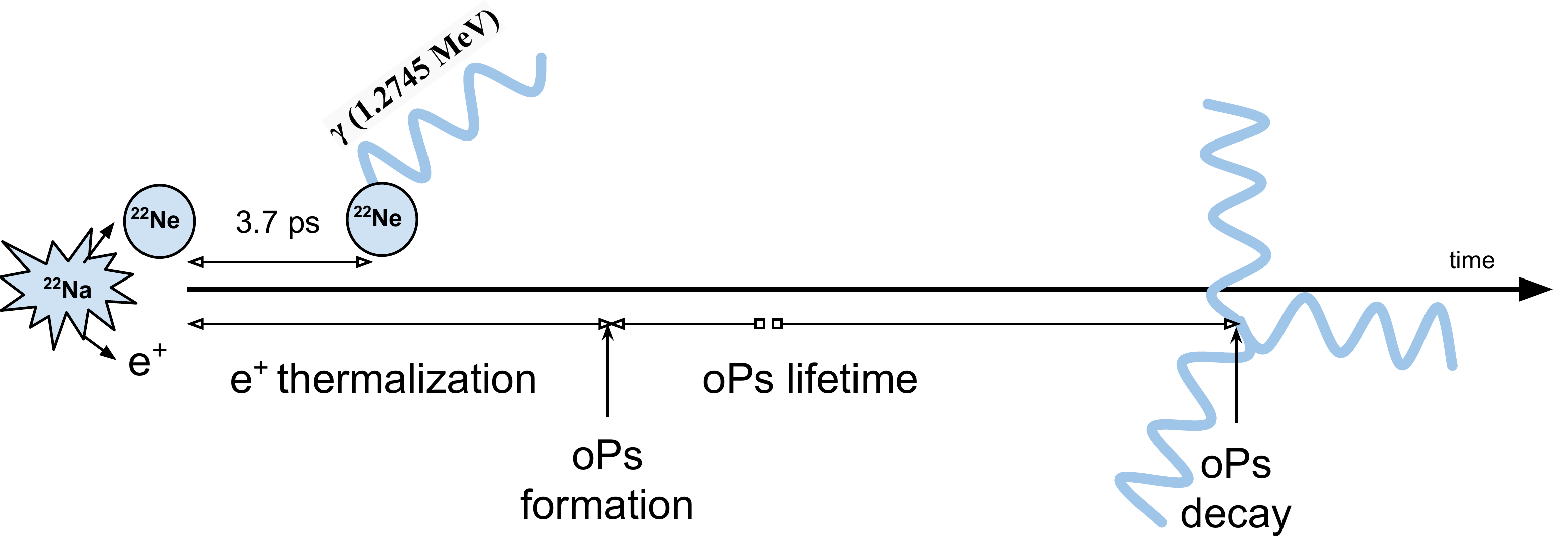}
    \vspace{-0.2cm}
    \caption{Scheme of sodium decay and formation of ortho-positronium.}
    \label{fig_positronium}
\end{figure*}

\begin{figure*}[p]
    \centering
    \vspace{-0.2cm}
    \includegraphics[width=0.8\columnwidth]{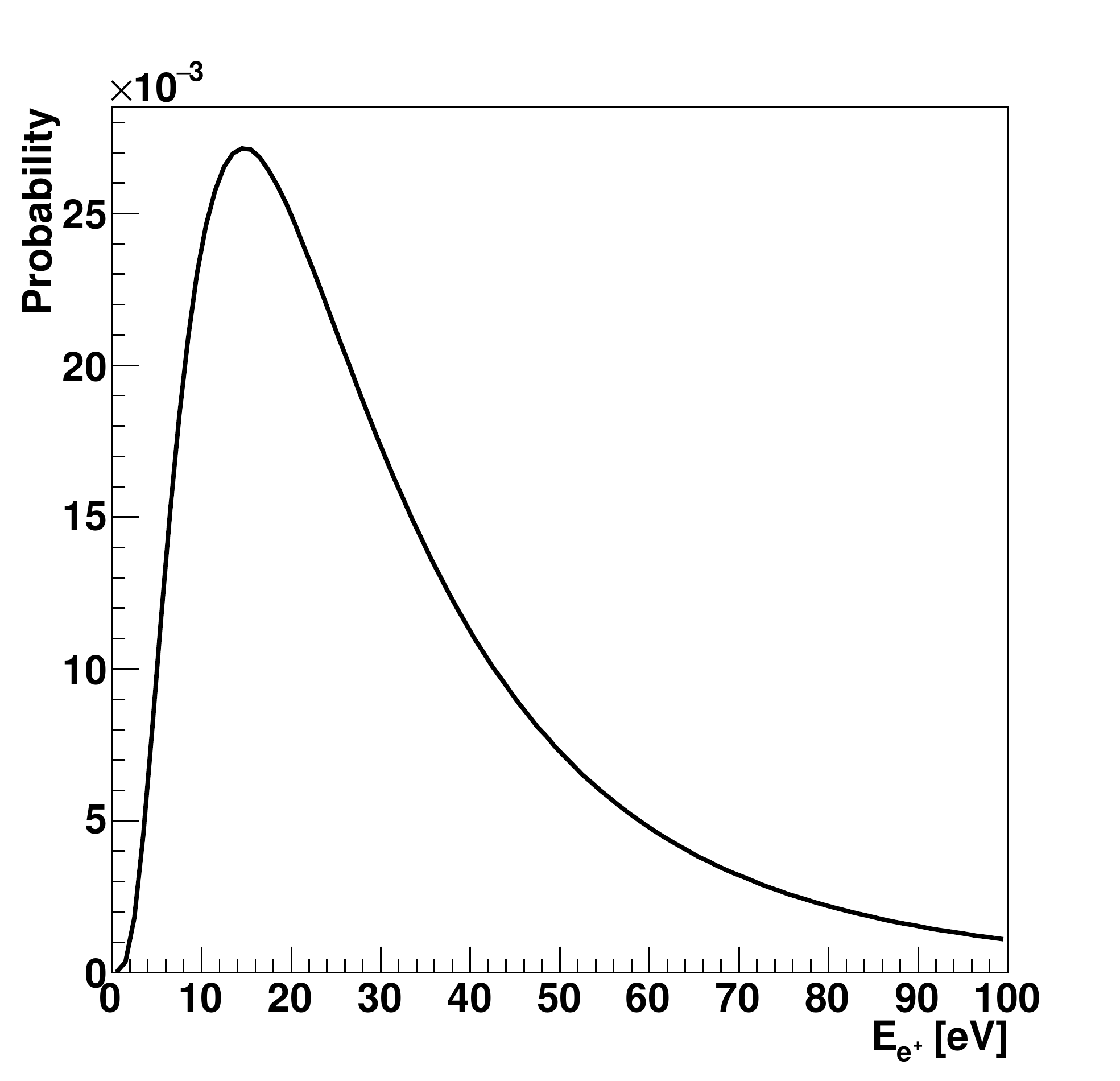}
    \vspace{-0.2cm}
    \caption{
        Simulated probability density function of positronium formation as a
        function of positron energy after thermalisation in the water.
    The distribution is adapted from reference~\cite{brazil}.}
    \label{pic_cross_section_positronium_probed}
\end{figure*}

\begin{figure*}[p]
    \centering
    \vspace{-0.5cm}
    \includegraphics[width=0.35\textwidth]{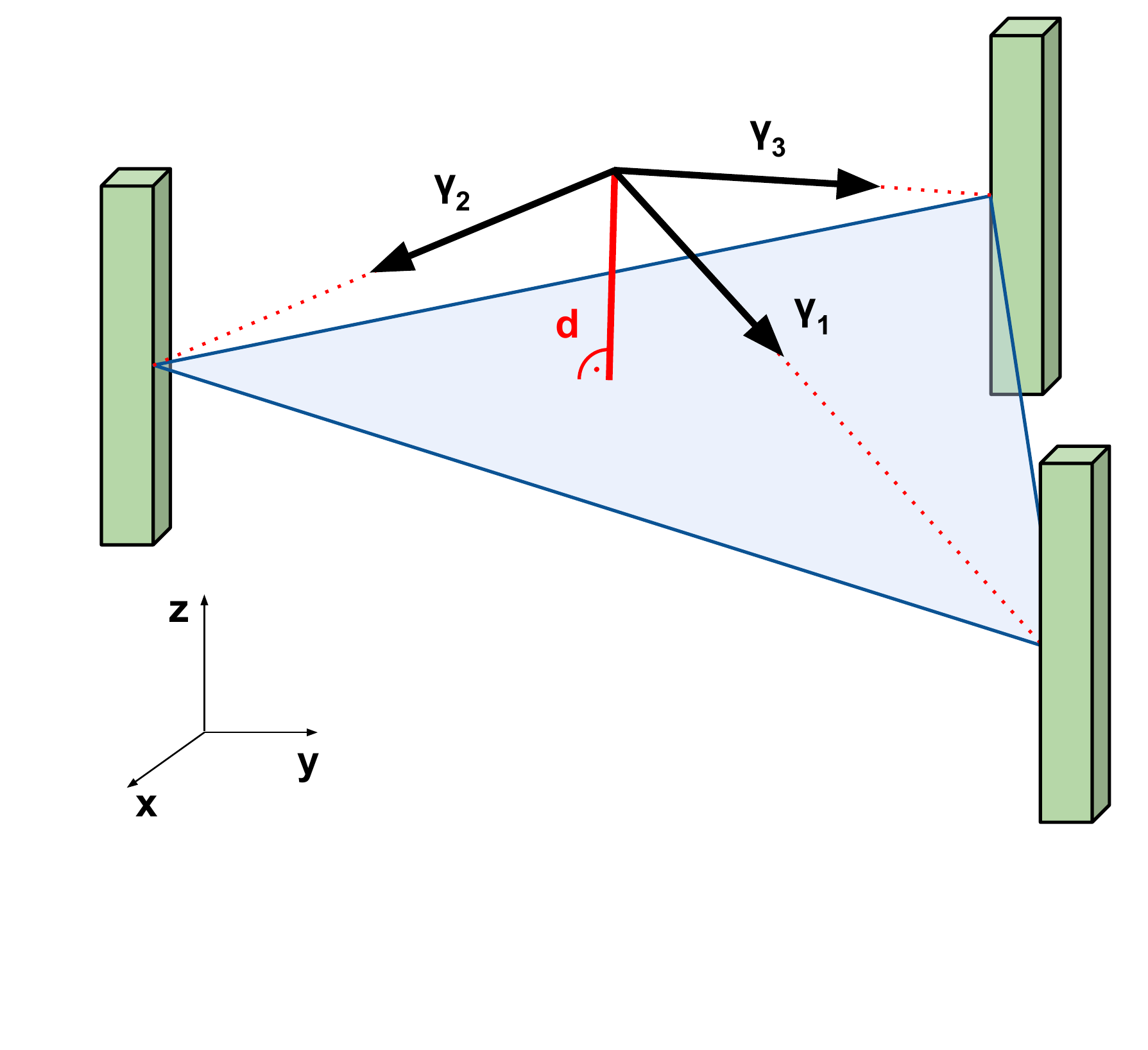}
    \includegraphics[width=0.4\textwidth]{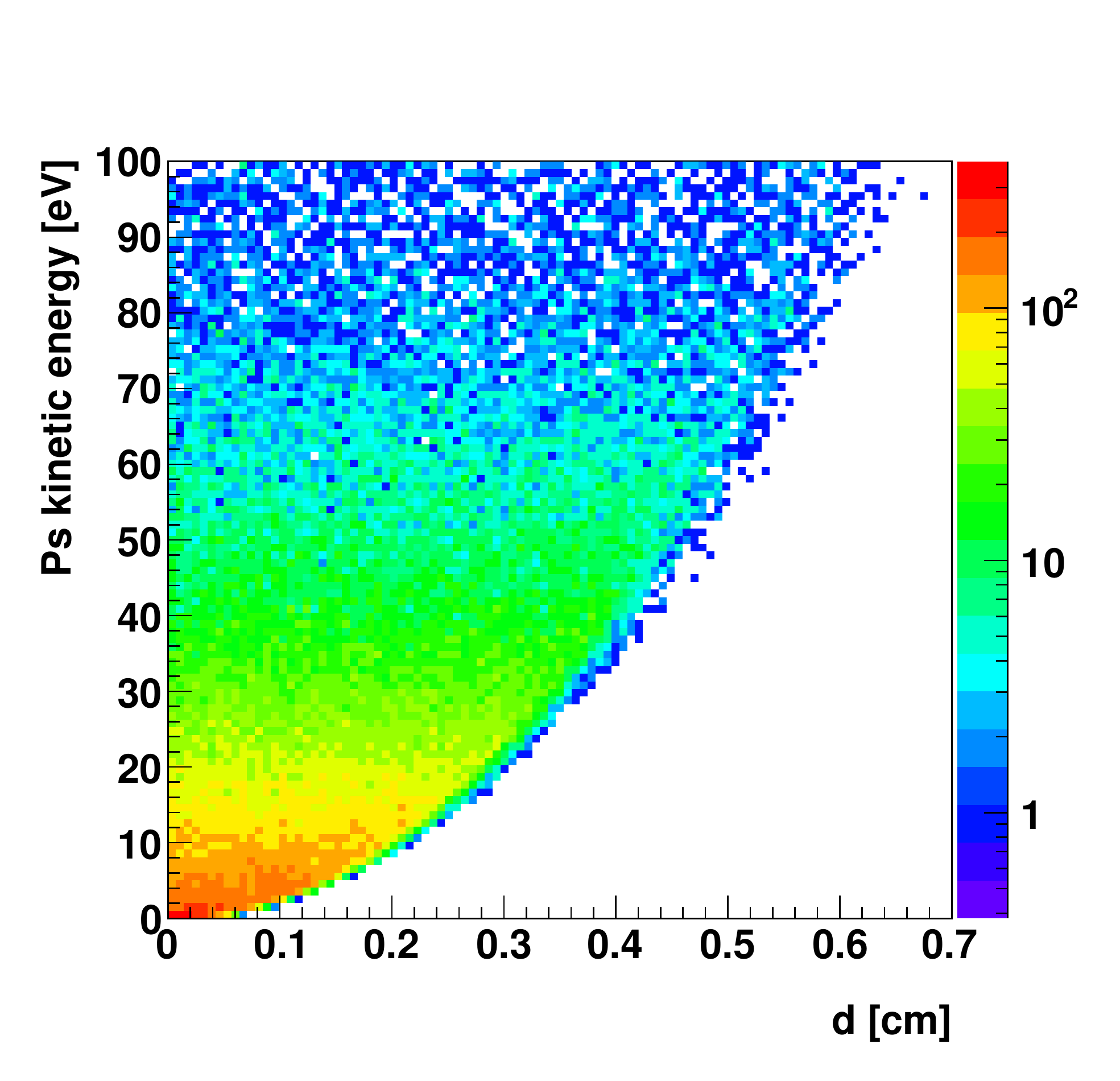}
    \vspace{-0.2cm}
    \caption{Left: Scheme of the ortho-positronium annihilation into three gamma quanta $\gamma_i$
        in the detector reference frame. Gamma quanta are not contained in a single plane due to
        non-zero kinetic energy of the ortho-positronium.
        In the experiment a plane of response can be determined from gamma quanta interaction position
        in the scintillators (green bars). The distance $d$ between plane of response and annihilation 
        vertex 
        gives information about annihilation position uncertainty.
        Right: Distribution of distance $d$ as  a function 
        of kinetic energy of ortho-positronium.
        Taking into account resolution of the J-PET  annihilation point reconstruction~\cite{Gajos:2016nfg}, the uncertainty 
        caused by o-Ps's boost is negligible.
    }
    \label{fig_boost}
\end{figure*}

In further considerations we will focus on sodium isotope, which is   commonly used as a source of positrons for various experiments and tests of detectors.
Pictorial representation of the studied  \ops process is shown in Figure~\ref{fig_positronium}.
In the conducted simulations we took into account the description of 
positron properties after thermalisation. Its energy was simulated
according to the distribution presented in Figure~\ref{pic_cross_section_positronium_probed}~\cite{brazil}.
The distribution of the initial positron kinetic energy depends only  on thermalisation processes.
This distribution  is taken into account in the transformation
of gamma quanta four-momenta from the rest frame of ortho-positronium to the laboratory frame.
In addition, the small distance traveled by positron in matter was taken into account.
Positron range depends on material properties and can be generated from profiles known in the literature~\cite{penelopet_positron_range}  provided by
many simulation packages, such as GATE~\cite{gate}
or PeneloPET~\cite{penelopet}. In this work the positron range distribution obtained by PeneloPET was 
adopted. 
Abovementioned effects introduce additional smearing of o-Ps annihilation position (see Figure~\ref{fig_boost}) and are included into performed simulations.

\subsection{\ops process}
\begin{figure*}[p]
    \centering
    \vspace{-0.3cm}
    \includegraphics[width=0.8\columnwidth]{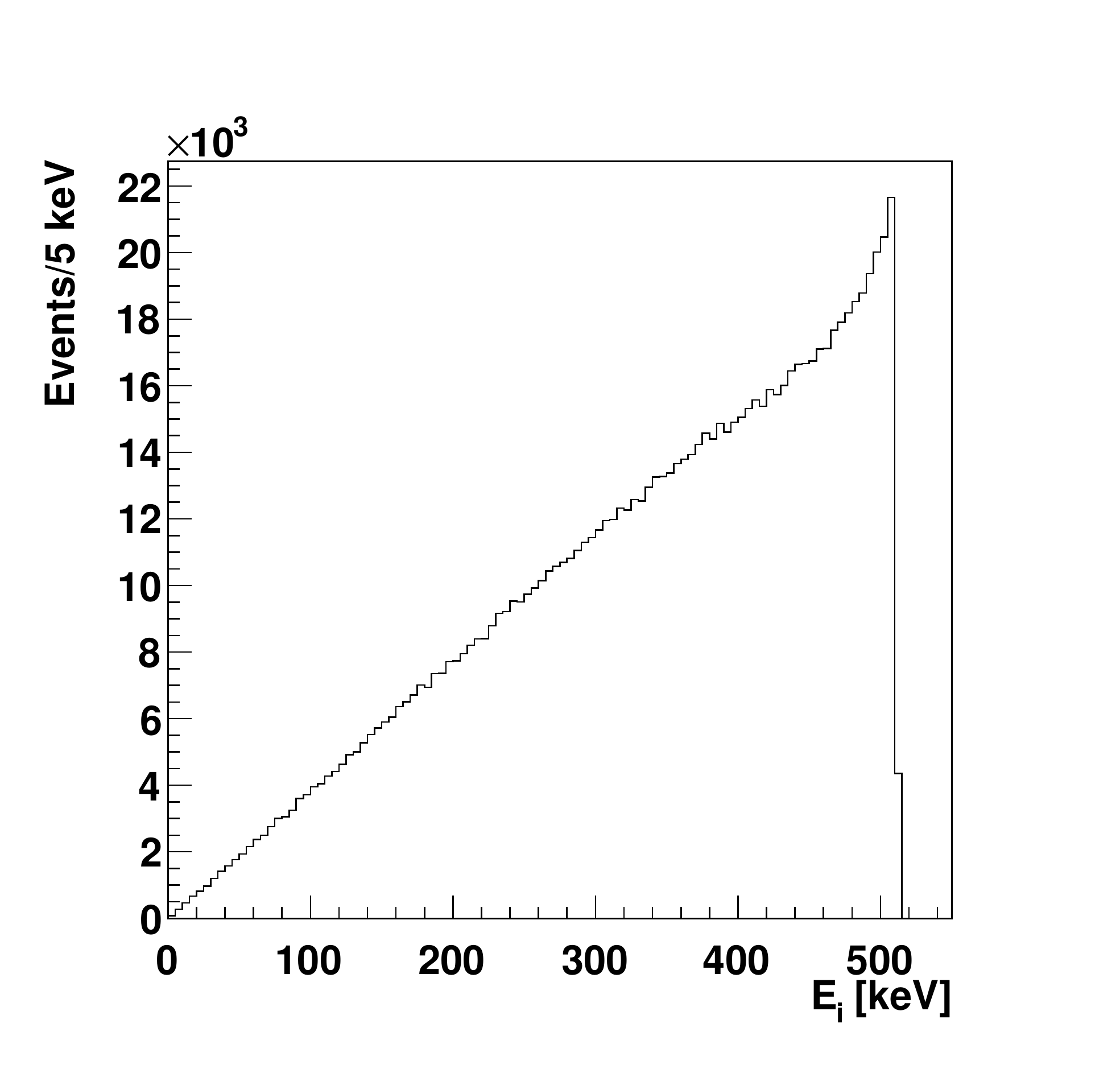}
    \vspace{-0.6cm}
    \caption{Energy spectrum of photons originating from three-photon
        annihilation of an electron and a positron.
    }
    \label{gamma_gen_ene}
\end{figure*}

\begin{figure*}[p]
    \centering
    \vspace{-0.4cm}
    \includegraphics[width=0.40\textwidth]{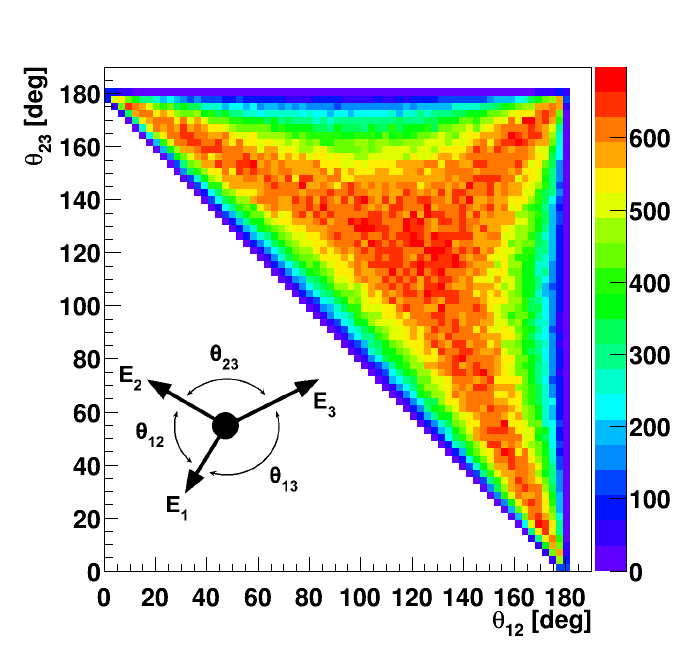}
    \includegraphics[width=0.40\textwidth]{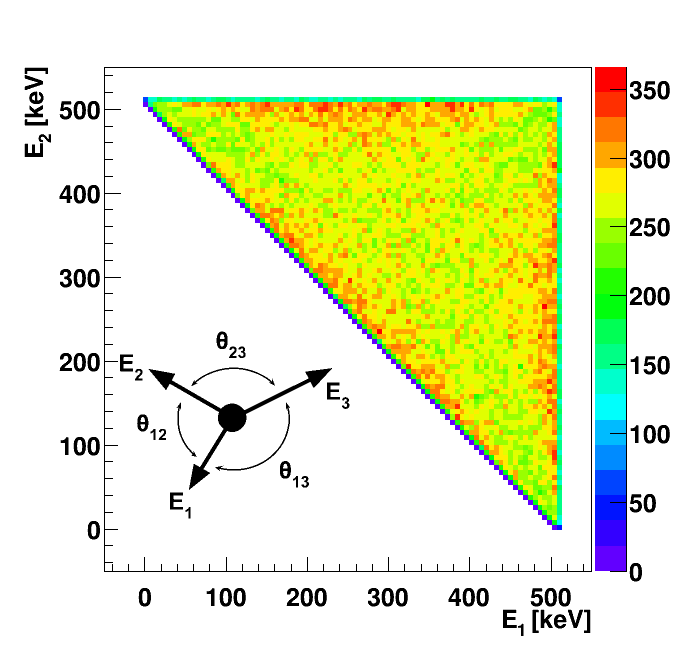}
    \vspace{-0.5cm}
    \caption{ 
        Distribution of angles (left) and Dalitz plot (right) of \ops annihilation. Boundaries are determined by kinematic constraints.
    }
    \vspace{-0.4cm}
    \label{fig_dalitz}
\end{figure*}

Positronium is the lightest purely leptonic system, and  it can annihilate only into 
gamma quanta. Those photons are coplanar in the Center of Mass (CM) frame due to the 
momentum conservation.
The cross-section for annihilation with formation of photons having frequencies $\omega_i$
can be expressed as~\cite{Lifshitz}:
\begin{align}
    \sigma_{3\gamma} & =  \frac{4 e^6}{v m_e^2}  \cdot \int_{0}^{m_e} \int_{m_e-\omega_{1}}^{m_e}
    \frac{\left( \omega_1 + \omega_2 - m_e \right)^2}{\omega_{1}^2 \omega_{2}^2} d \omega_1 d \omega_2
    \nonumber \\
    & = \frac{4 e^6}{v m_e^2} \cdot \frac{\pi^2 - 9}{3}
    \label{cross_section}
\end{align}
where 
$m_e$ is electron mass,
$v$ denotes electron-positron relative velocity, 
$e$ is the elementary charge.
In above formula the conservation of 4-momentum allows to eliminate one of the frequencies ($\omega_{3}$).
Equation~\ref{cross_section} 
results in the characteristic  energy distribution 
of gamma quanta (see Figures~\ref{gamma_gen_ene} and~\ref{fig_dalitz}).
\vspace{0.3cm}
\section{Simulated geometries}
\label{sec:MC_det}
\vspace{0.6cm}
A few possible detector geometries were simulated. They are referred to as
J-PET, J-PET+1, J-PET+2 and J-PET-full:
\begin{description}[style=multiline,leftmargin=2.5cm]
    \item[J-PET] corresponds to the already built detector~\cite{Moskal:2016moj} with 3 layers of  scintillators 
        (from in to out: 48+48+96 scintillators).   
    \item[J-PET+1] the J-PET geometry extended by an additional layer 
        filled by 96 scintillators.
    \item[J-PET+2]  geometry assumes complete fulfillment of all available layers in the J-PET detector
        (48+48+96+96+96).
    \item[J-PET-full] detector with fully coverage of four plastic scintillator layers.
\end{description}
\tikzset{new spy style/.style={spy scope={%
            size=1.25cm, 
            connect spies,
            every spy on node/.style={
                rectangle,
                draw,
                orange, 
            },
            every spy in node/.style={
                draw,
                rectangle,
                orange, 
            }
        }
    }
} 
\begin{figure*}[p]
    \begin{subfigure}[b]{0.5\textwidth}
        \begin{tikzpicture}[new spy style]
            \node  { \includegraphics[width=0.5\textwidth]{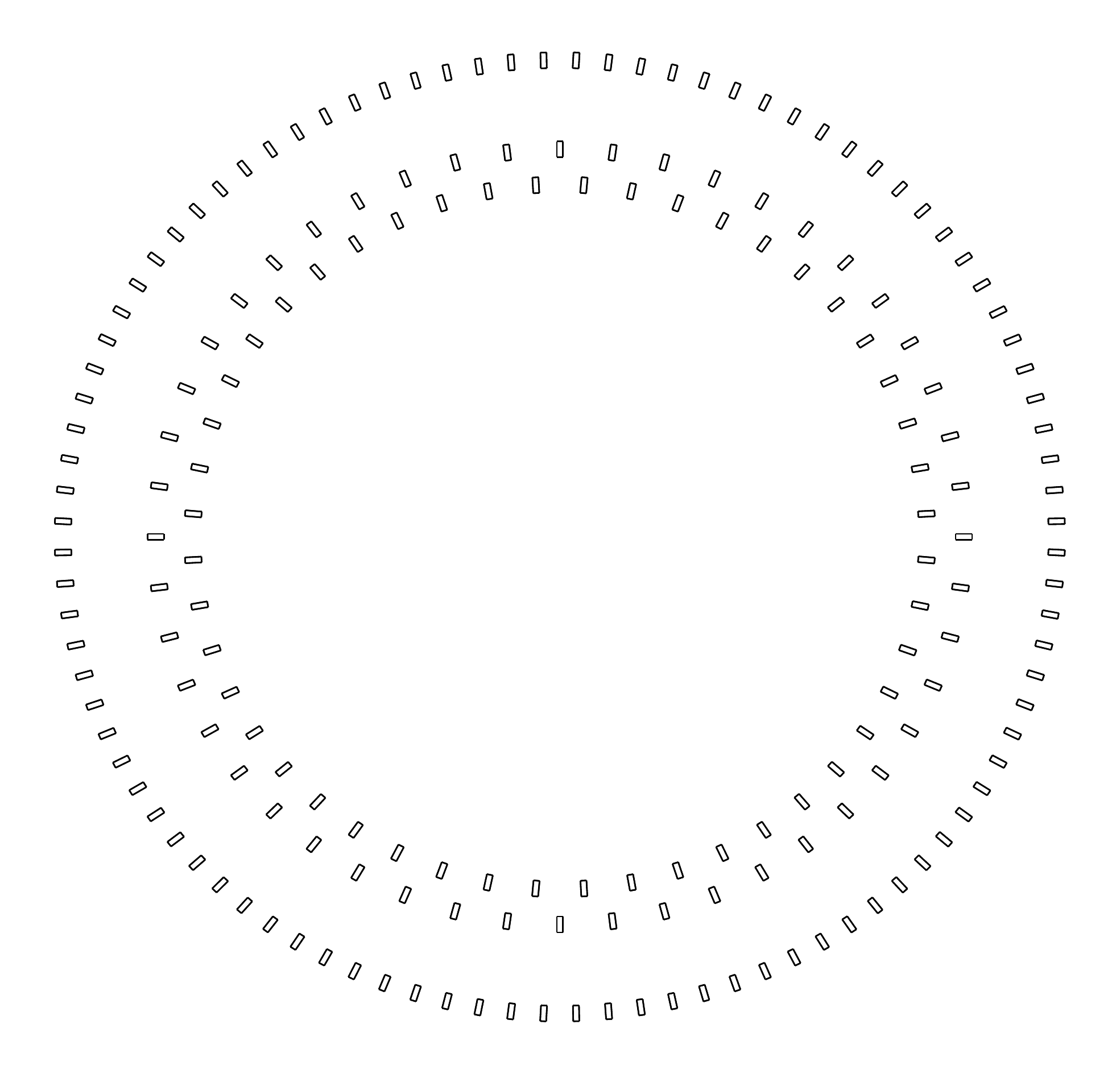} };
            \node[draw=none] at (0.1,0.1) {\bf J-PET};
            \spy[width=2cm,height=3cm,magnification=2.5] on (1.7,-0.15) in node [fill=white] at (3.5,0);  
            \draw [<->,thick] (-2,-1) node (yaxis) [left] {$y$}
            |- (-1,-2) node (xaxis) [below] {$x$};
        \end{tikzpicture}
    \end{subfigure}
    \begin{subfigure}[b]{0.5\textwidth}
        \begin{tikzpicture}[new spy style]
            \node  { \includegraphics[width=0.5\textwidth]{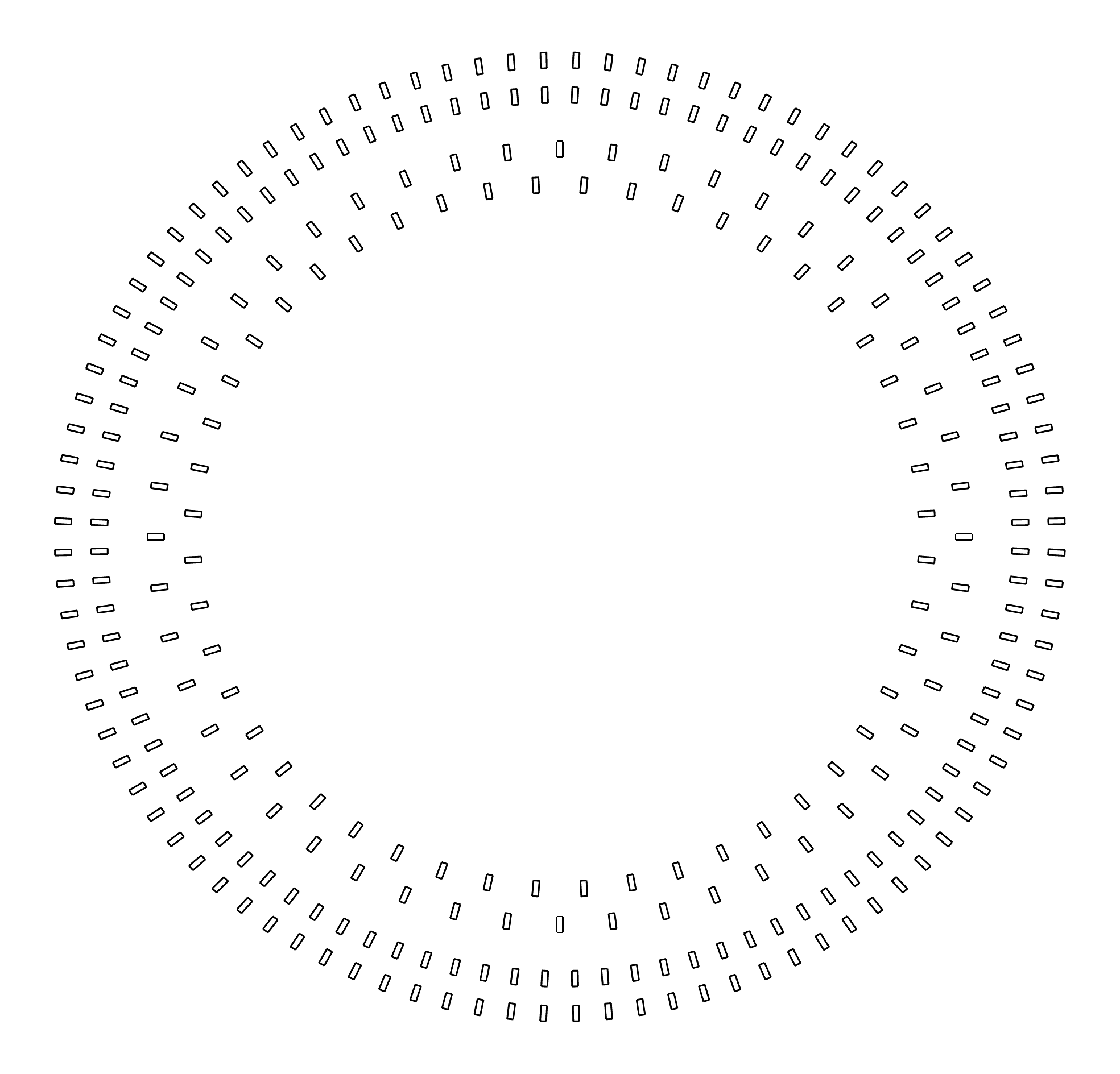} };
            \node[draw=none] at (0.1,0.1) {\bf J-PET+1};
            \spy[width=2cm,height=3cm,magnification=2.5] on (1.7,-0.15) in node [fill=white] at (3.5,0);  
            \draw [<->,thick] (-2,-1) node (yaxis) [left] {$y$}
            |- (-1,-2) node (xaxis) [below] {$x$};
        \end{tikzpicture}
    \end{subfigure}
    \newline 
    \begin{subfigure}[b]{0.5\textwidth}
        \begin{tikzpicture}[new spy style]
            \node  { \includegraphics[width=0.5\textwidth]{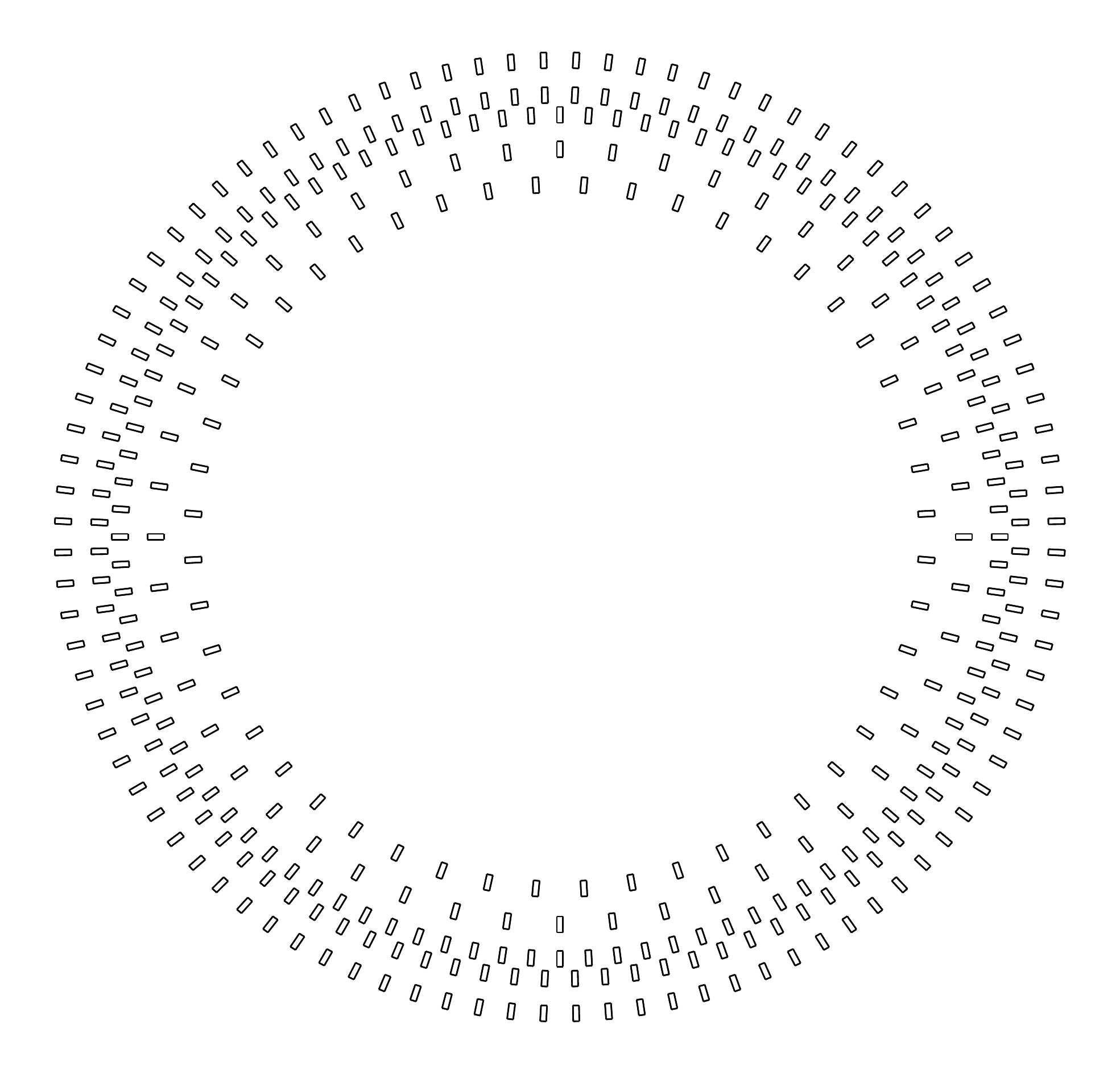} };
            \node[draw=none] at (0.1,0.1) {\bf J-PET+2};
            \spy[width=2cm,height=3cm,magnification=2.5] on (1.7,-0.15) in node [fill=white] at (3.5,0);  
            \draw [<->,thick] (-2,-1) node (yaxis) [left] {$y$}
            |- (-1,-2) node (xaxis) [below] {$x$};
        \end{tikzpicture}
    \end{subfigure}
    \begin{subfigure}[b]{0.5\textwidth}
        \begin{tikzpicture}[new spy style]
            \node  { \includegraphics[width=0.5\textwidth]{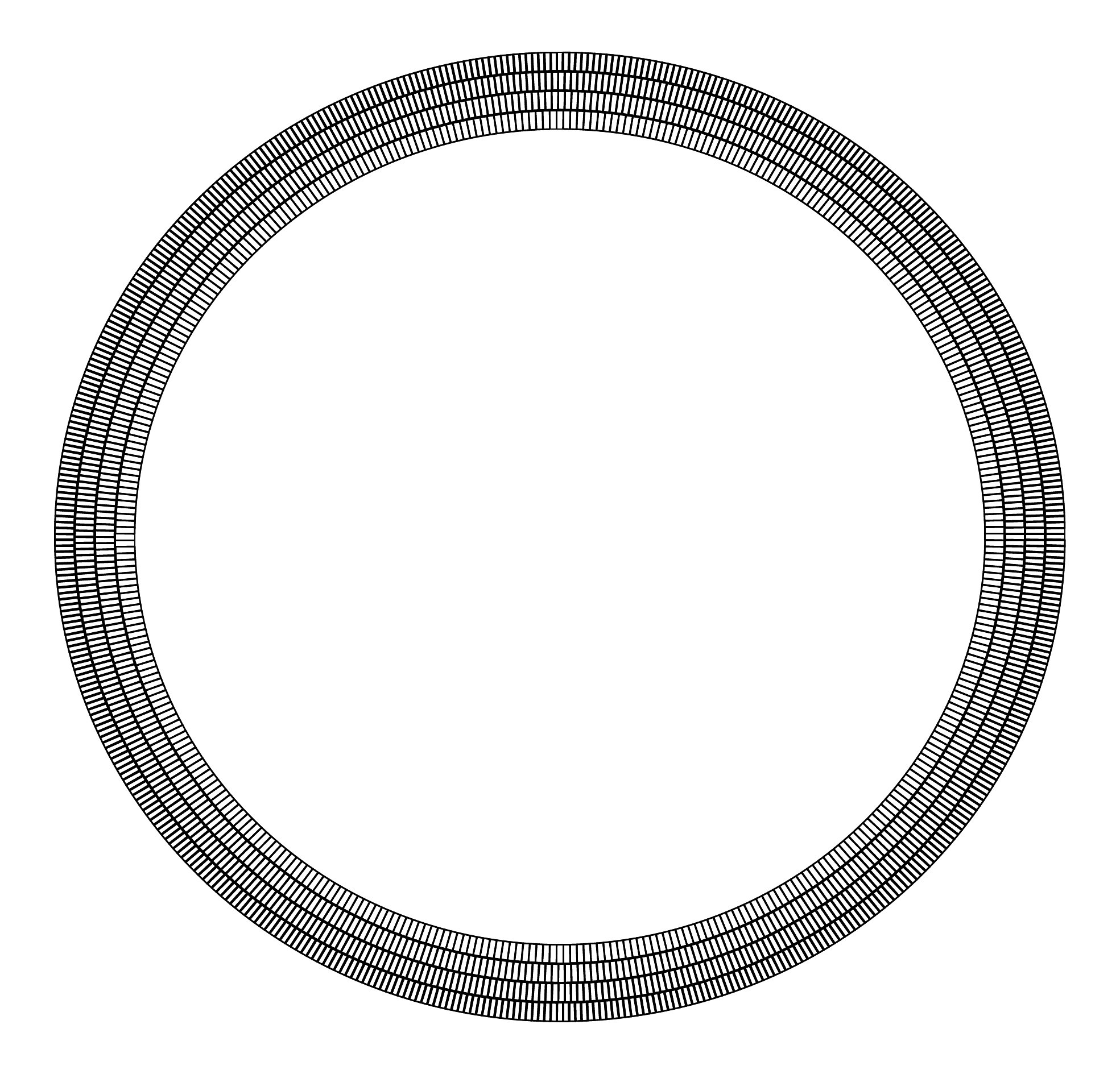} };
            \node[draw=none] at (0.1,0.1) {\bf J-PET-full};
            \spy[width=2cm,height=3cm,magnification=2.5] on (1.7,-0.15) in node [fill=white] at (3.5,0);  
            \draw [<->,thick] (-2,-1) node (yaxis) [left] {$y$}
            |- (-1,-2) node (xaxis) [below] {$x$};
        \end{tikzpicture}
    \end{subfigure}
    \vspace{-0.7cm}
    \caption{Transverse view of simulated geometries.} 
    \label{fig_xy_view_of_real_detector}
\end{figure*}
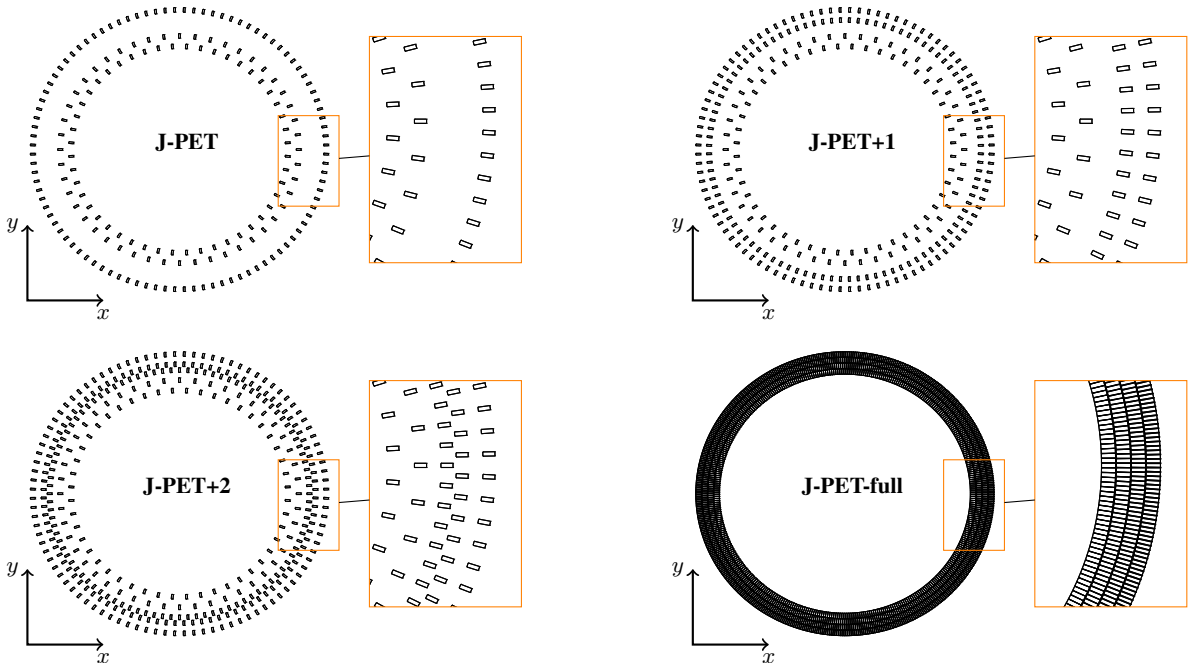
The details of the simulated geometries are presented in Figure~\ref{fig_xy_view_of_real_detector}
and Table~\ref{tab_real_detector}.
Comparison of the results for all above mentioned options
shows the  accuracy of \ops registration achievable at 
current J-PET setup and upgrades planned
\noindent
in the next two years, as well as for the J-PET-full detector.
In all the cases simulations assume the usage of   EJ-230 plastic scintillator strips 
(dimensions $1.9\mbox{ cm} \times  0.7\mbox{ cm} \times 50.0\mbox{ cm}$), with the longest side of the 
scintillator arranged along the $z$~axis.
\begin{table*}[p]
    \centering
    \caption{
        Details of simulated layers of the J-PET geometry.
        J-PET detector has been already built~\cite{Moskal:2016moj}. 
        The mechanical construction for the next phases J-PET+1 and J-PET+2 
        is also prepared and the hardware upgrade is planned within the next two years.
    }
    \label{tab_real_detector}
    \vspace{-0.3cm}
    \begin{tabular*}{0.49\textwidth}{@{\extracolsep{\fill}}|C{0.07\textwidth}|C{0.1\textwidth}|C{0.07\textwidth}|C{0.15\textwidth}|@{}}
        \hline
            Layer number & Layer radius with respect to the center of scintillator & Number of scintillators in the layer 
            & Angular displacement of $n_i$ scintillator  \\ \hline \hline
            \multicolumn{4}{c}{\bf J-PET} \\ \hline
            1 & 42.50 cm & 48 & $n_i \times 7.5^{\circ}$ \\ \hline 
            2 & 46.75 cm & 48 & $n_i \times 7.5^{\circ} + 3.75^{\circ}$ \\ \hline
            3 & 57.50 cm & 96 & $n_i \times 3.75^{\circ} + 1.875^{\circ}$ \\ \hline 
            \multicolumn{4}{c}{\bf J-PET+2} \\ \hline
            1 & 42.50 cm & 48 & $n_i \times 7.5^{\circ}$ \\ \hline 
            2 & 46.75 cm & 48 & $n_i \times 7.5^{\circ} + 3.75^{\circ}$ \\ \hline
            3 & 50.90 cm & 96 & $n_i \times 3.75^{\circ}$\\ \hline
            4 & 53.30 cm & 96 & $n_i \times 3.75^{\circ} + 1.875^{\circ}$ \\ \hline
            5 & 57.50 cm & 96 & $n_i \times 3.75^{\circ} + 1.875^{\circ}$ \\ \hline 
        \end{tabular*}
    \begin{tabular*}{0.49\textwidth}{@{\extracolsep{\fill}}|C{0.07\textwidth}|C{0.1\textwidth}|C{0.07\textwidth}|C{0.15\textwidth}|@{}}
        \hline
            Layer number & Layer radius with respect to the center of scintillator & Number of scintillators in the layer 
            & Angular displacement of $n_i$ scintillator  \\ \hline \hline
            \multicolumn{4}{c}{\bf J-PET+1} \\ \hline
            1 & 42.50 cm & 48 & $n_i \times 7.5^{\circ}$ \\ \hline 
            2 & 46.75 cm & 48 & $n_i \times 7.5^{\circ} + 3.75^{\circ}$ \\ \hline
            3 & 53.30 cm & 96 & $n_i \times 3.75^{\circ}+1.875^{\circ} $ \\ \hline
            4 & 57.50 cm & 96 & $n_i \times 3.75^{\circ}+1.875^{\circ}$ \\ \hline 
            \multicolumn{4}{c}{\bf J-PET-full} \\ \hline
            1 & 43.0 cm & 400 & $n_i \times 0.9^{\circ}$  \\ \hline
            2 & 45.0 cm & 437 & $n_i \times 0.82^{\circ}$\\ \hline
            3 & 47.0 cm & 473 & $n_i \times 0.76^{\circ}$\\ \hline
            4 & 49.0 cm & 508 & $n_i \times 0.71^{\circ}$\\ \hline
        \end{tabular*}
\end{table*}
\begin{figure*}[p]
    \centering
    \vspace{-1.2cm}
    \includegraphics[width=0.76\columnwidth]{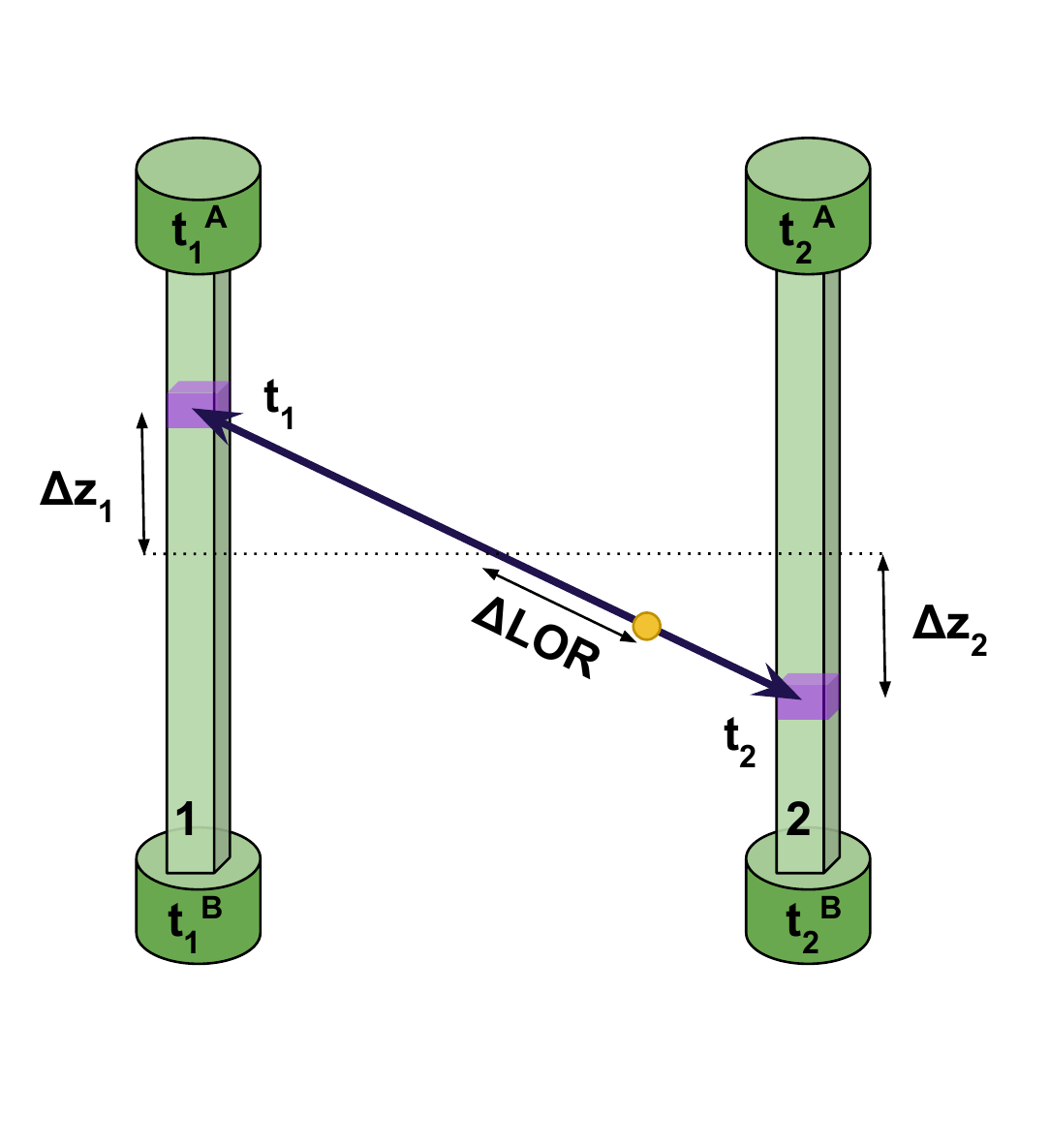}
    \vspace{-1.3cm}
    \caption{ Registration of the 
        signals arrival time on the two ends of a single scintillator ($t_1^A$, $t_1^B$
        and $t_2^A$, $t_2^B$ for the first and second strip, respectively) allows to
        determine the distance from the scintillators centers ($\Delta z_{1,2}$)
        and times ($t_{1,2}$) when gamma quanta interacts with scintillators. 
        Then the Line of Response can be determined as well as the
        displacement of the annihilation position from its center ($\Delta \mbox{LOR}$).
    }
    \label{two_strip}
\end{figure*}
\begin{figure*}[p]
    \centering
    \vspace{-0.5cm}
    \includegraphics[width=0.38\textwidth]{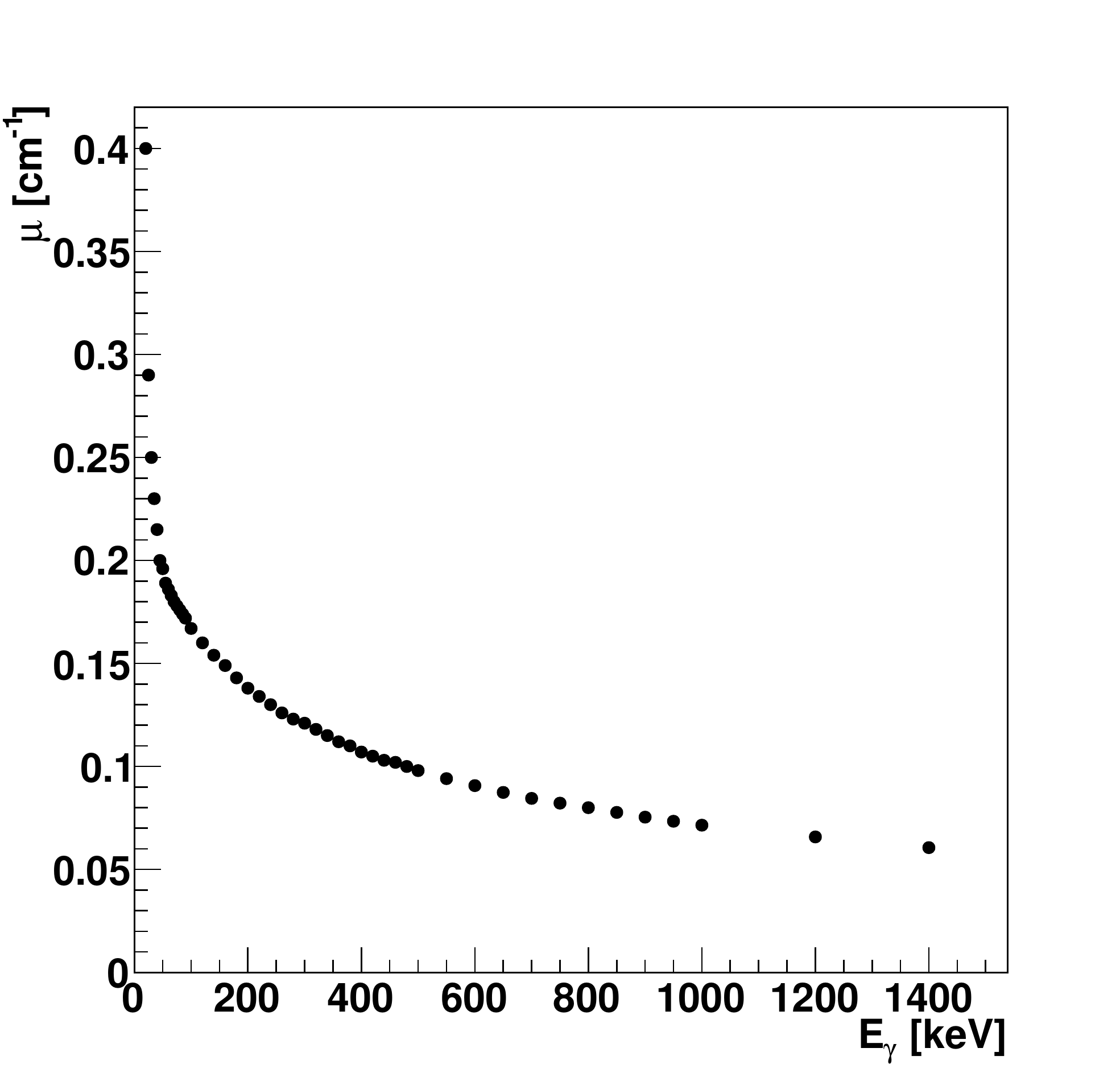}
    \includegraphics[width=0.38\textwidth]{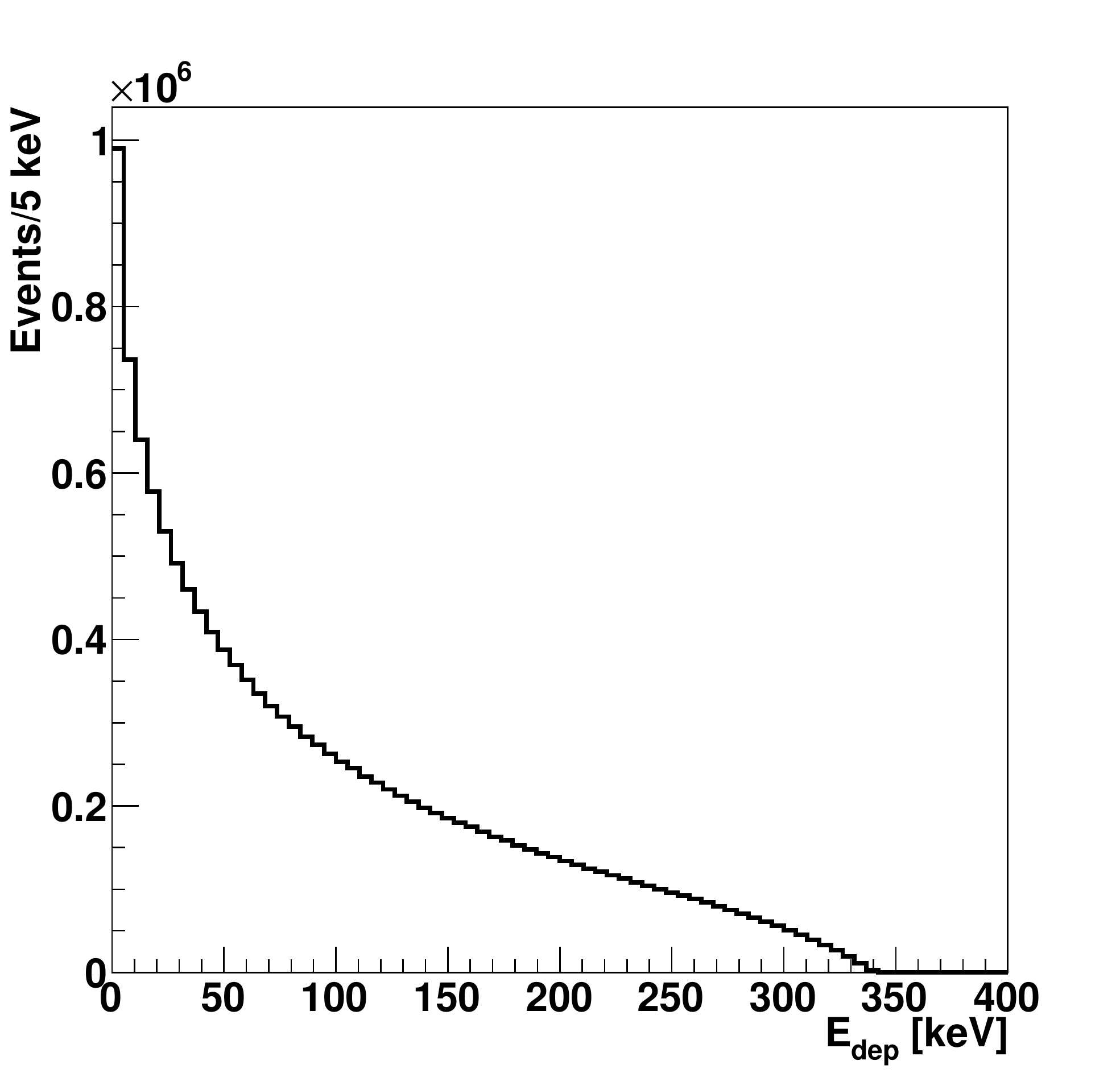}
    \vspace{-0.4cm}
    \caption{
        Left: Dependency of attenuation coefficient on incident  gamma quanta energy.
        Data taken from~\cite{SaintGobainCrystals}.
        Right: 
        Distribution of energy deposited by gamma quanta in plastic scintillators
        originating from \ops annihilations.
        The shown spectrum is a convolution of the energy distribution of gamma quanta from the 
        \ops  decay (Figure~\ref{gamma_gen_ene})
        and the Klein-Nishina distribution of kinetic energy of electrons acquired via
        Compton scattering~\cite{radiation_interaction}.   
        Spectrum includes the absorption dependence 
        on the energy (left panel) and the detector energy resolution.
    }
    \label{fig_ene_depos}
\end{figure*}
\section{J-PET detector properties}
\label{sec:det}
\vspace{0.5cm}
The  multipurpose detector (J-PET) constructed at the Jagiellonian University
of which novelty lies in using large blocks of plastic scintillators 	
instead of crystals as detectors of annihilation quanta,
requires the usage of the time of signals, instead of their amplitude, and allows
to obtain time resolution better than 100~ps~\cite{Moskal:2014sra}.
\subsection{Determination of hit and time position at J-PET}
\vspace{0.3cm}
Reconstruction of time and gamma quanta hit position in $i^{th}$ plastic scintillator 
can be based on the time values ($t_i^A$, $t_i^B$) of scintillation light registration in photomultipliers
located at the ends of single plastic scintillator strip. Then the distance $(\Delta z_i)$ 
along the strip between its center and the hit position can be expressed~as:
\begin{equation}
    \Delta z_i = \frac{(t_i^A - t_i^B)\cdot v}{2},
\end{equation}
where $v$ is the light velocity in the plastic scintillator.
Based on this information, in case of two-gamma quanta annihilation, the Line of
Response (LOR) and the annihilation position along it can be determined (see Figure~\ref{two_strip}). 

In case of three-gamma annihilation, the registered gamma quanta are coplanar
(o-Ps kinetic energy can be neglected, see  section~\ref{sec:formation}) and the 
registered hit-points form the Plane-of-Response (POR)
(see Figure~\ref{scheme_detector}, right panel).  
In this case the annihilation position can be determined using the novel 
reconstruction based on trilateration method (see Section~\ref{sec:rec_algo}).
The obtained energy and time resolution of registered gamma quanta were experimentally determined 
and within the range of deposited energy ($E_{dep}$)  $\in (200,340) \mbox{ keV}$, are equal~to~\cite{Moskal:2014sra}:
\begin{align}
    \sigma( T^0_{hit} ) & \approx 80\mbox{ ps} \label{sigma_t_hit}, 
     \\
     \frac{\sigma(E)}{E} & = \frac{0.44}{\sqrt{ E \mbox{ [MeV]} }}. 
    \label{sigma_E} 
\end{align}
For lower energies the time resolution can be expressed~as a function of
deposited energy~($E_{dep}$):
\begin{equation}
    \sigma( T_{hit} (E_{dep})) = \frac{ \sigma(T^0_{hit}) \mbox{[ps]} }{ \sqrt{\frac{E_{dep} \mbox{[keV]}}{270}} }.
    \label{sigma_E_lower_energy}
\end{equation}
Considering the most challenging time reconstruction for gamma quanta with low energies (around 50 keV), 
one can see that the J-PET detector provides a precision
on the level of two hundred picoseconds.
In the commercial PET systems the events with an energy deposition lower than about 400~keV~\cite{bettinardi,surti} are discarded.
\subsection{Spectra of deposited energy}
\label{sec:specta_energy}
\vspace{0.3cm}
The probability of incident  gamma quanta registration is a function of the attenuation coefficient $\mu$ and distance 
that gamma quantum travels through the material. In the simulations the attenuation coefficient was parametrized as a function
of incident gamma quanta energy (see Figure~\ref{fig_ene_depos}, left panel).

\begin{figure}
    \centering
    \includegraphics[width=0.45\textwidth]{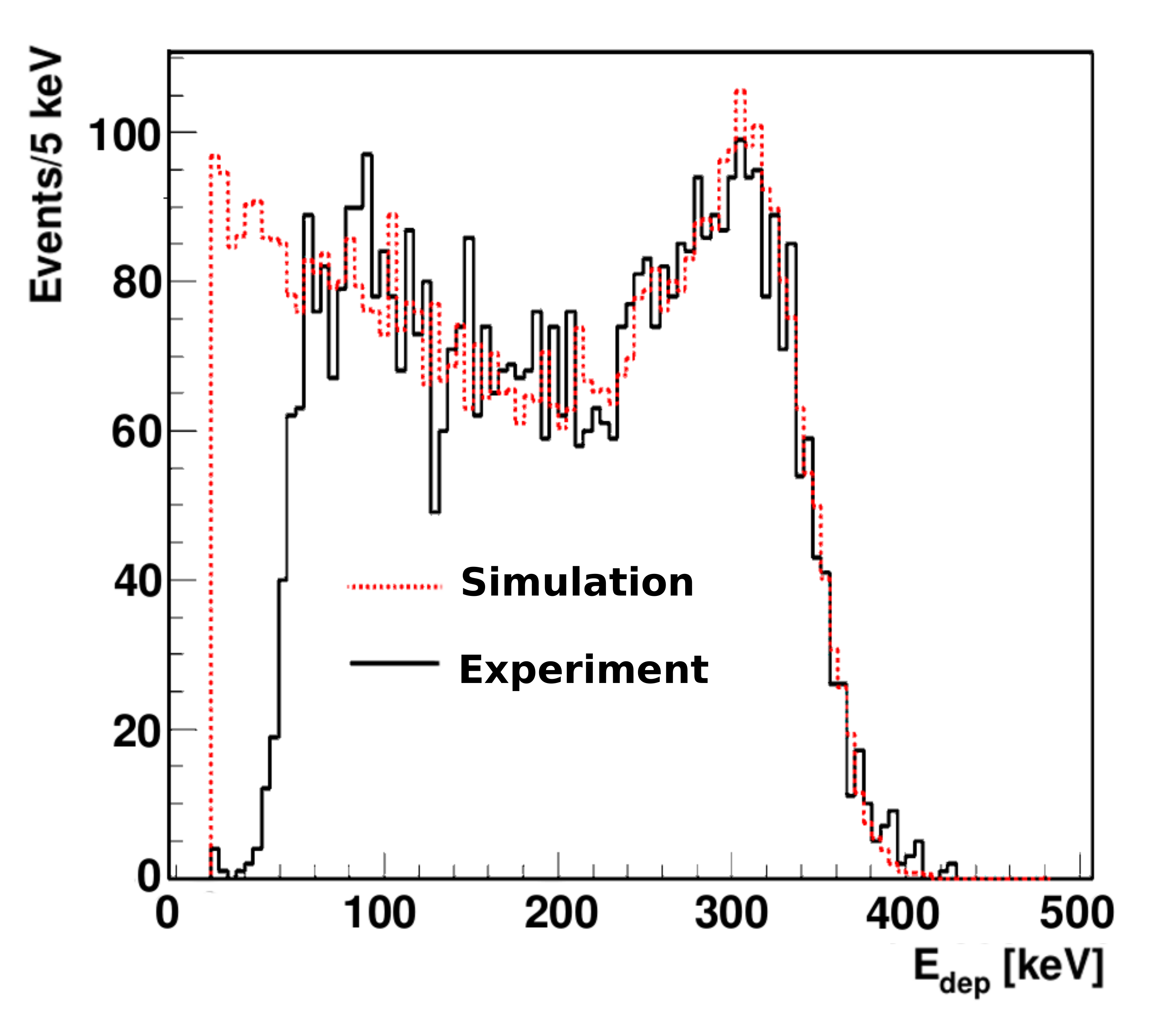}
    \caption{
        Spectra of simulated (red, dashed line) and measured energy (solid, black line) deposition by 511~keV gamma quanta in J-PET detector.
        The simulated spectrum was normalized to the experimental one, and simulations were performed taking into account 
        the energy resolution (Equation~\ref{sigma_E}).  The left part of the experimental spectrum was cut due to the triggering threshold applied in the experiment.
    }
    \label{fig_exp_ve_theory}
\end{figure}

Gamma quanta interact with plastic scintillators mainly via Compton effect and 
the characteristic spectra 
of deposited energy are described by Klein-Nishina formula~\cite{radiation_interaction,Klein}. 
The distribution for 511~keV incident gamma quantum 
is shown in Figure~\ref{fig_exp_ve_theory}.
Energy of single gamma quanta from ortho-positronium annihilation is within $[0,511]$~keV energy range 
and the spectrum of deposited energy via Compton effect for the corresponding energy range
is presented in Figure~\ref{fig_ene_depos} (right panel).
\subsection{Background rejection}
\label{sec:bck_rejection}
\vspace{0.3cm}
Direct annihilation of positron with electron, as well as intrinsic annihilation of 
para-positronium, are both characterized by short times of $\sim$400~ps and $\sim 125$~ps, respectively.
For comparison, an ortho-positronium lifetime in vacuum amounts to about 142~ns~\cite{Ramadhan,Vallery,Jinnouchi}. 
Therefore, events corresponding to direct annihilation and decay of para-positronium can be reduced to a negligible level by 
requiring the time difference between de-excitation photon and annihilation 
photons detection to be larger than e.g. 20~ns. 
However, such lifetime criterion cannot discriminate pick-off and conversion processes of o-Ps 
which may lead to the annihilation into $2\gamma$ quanta.  

\begin{figure*}
    \centering
    \includegraphics[width=0.8\textwidth]{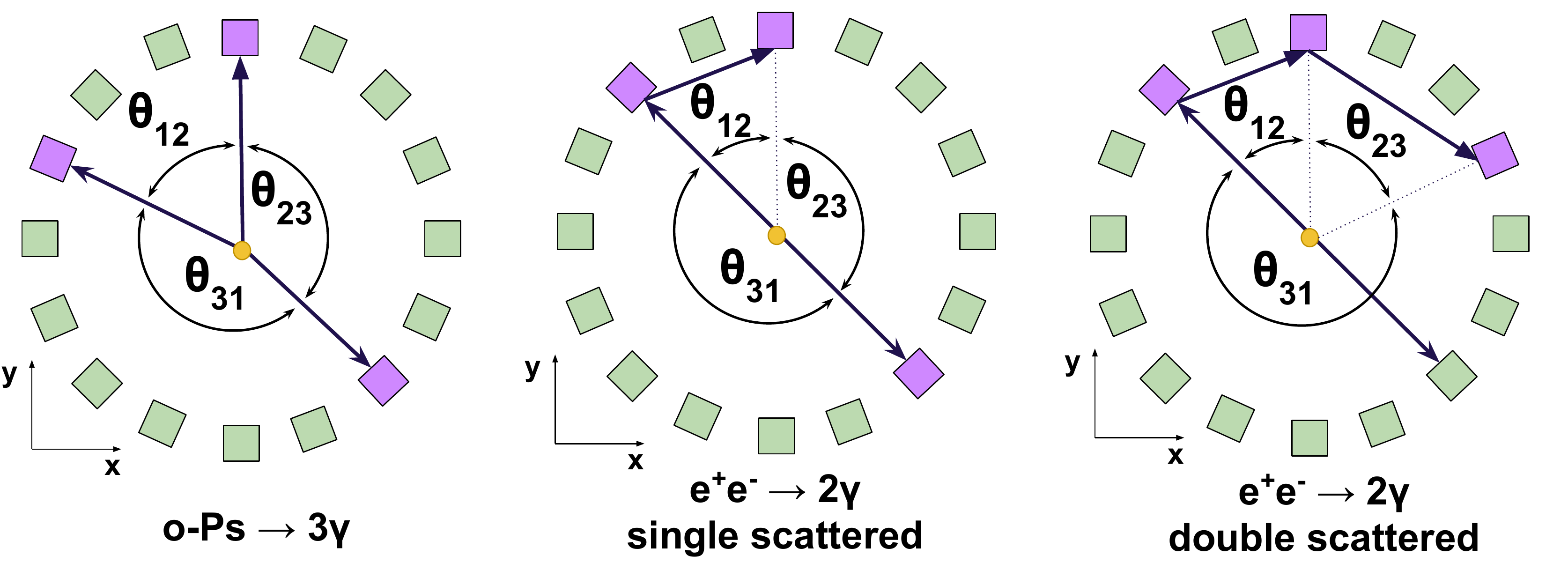} 
    \caption{Pictorial illustration of the possible response of the detector to \ops and $e^+e^-$ annihilation into $2\gamma$.
        Arranged circularly  squares represents scintillator strips - purple and green  colors indicate strips where the
        gamma quanta were or were not registered, respectively. The arrows represents gamma quanta occurring in the events, 
        while dotted lines indicate naively reconstructed gamma quanta. 
    Examples of primary and secondary scatterings are depicted.}
    \label{fig_scattering}
\end{figure*}

\begin{figure}
    \centering
    \includegraphics[width=0.5\textwidth]{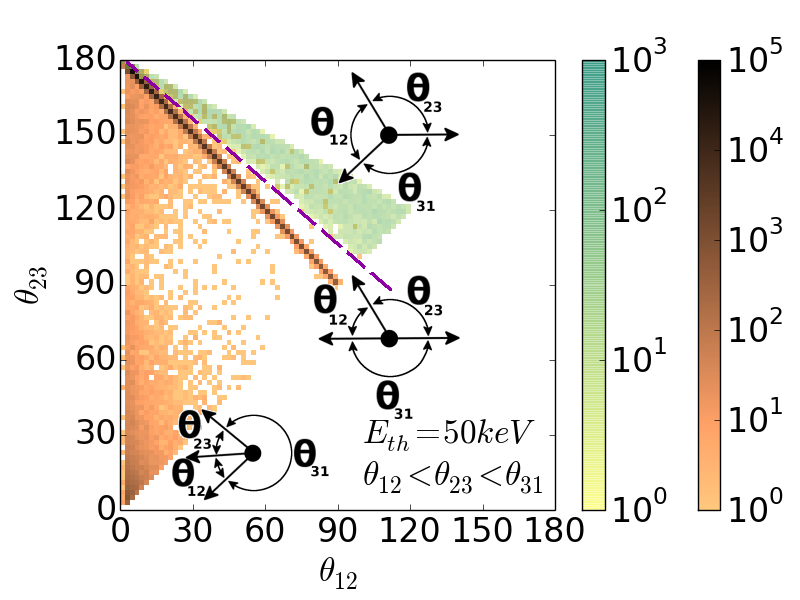} 
    \caption{Distribution of \ops (green) and scattered 
        events (brown) as a function of $\theta_{12}$ vs $\theta_{23}$ angles.
        Events, where one of the gamma from $e^+e^- \rightarrow 2 \gamma$ annihilation is registered in the detector
        while the other is scattered and cause signals in two detectors, lies on the diagonal of the plot.
        Events where one gamma is missing detection, and the other undergoes two scatterings
        are localized below the diagonal line.
        Example of analysis cut, rejecting $3\%$ of signal and reducing background by factor 
        $10^{4}$, is shown as a dashed purple line.
        Distribution includes the angular resolution of the J-PET detector.
    }
    \label{fig_theta_cut}
\end{figure}

Annihilation into $2\gamma$ may mimic a registration of $3\gamma$ annihilation 
due to the secondary scatterings in the detector.
Such scattering is shown pictorially in  Figure~\ref{fig_scattering}.
For the reduction of this background the following complementary methods can be considered,  based on information~of:
\begin{itemize}
    \item relation between position of the individual detectors and the time difference
        between registered hits,
    \item  angular correlation of relative angles between the  gamma quanta propagation directions,
    \item the distance between the origin of the annihilation (position of the annihilation chamber) and the decay plane.
\end{itemize}  
In Figure~\ref{fig_theta_cut} we show as an example  spectra  the $\theta_{23}$ vs $\theta_{12}$ distribution, where $\theta_{ij}$ 
are the ordered opening angles ($\theta_{12} < \theta_{23} < \theta_{13}$) between registered gammas.
For the \ops process, due to the momentum conservation, $\theta_{23} > 180^{\circ} - \theta_{12}$   and therefore events corresponding to the \ops  decay 
will lie above the diagonal, as shown in green colour in Figure~\ref{fig_theta_cut}.
Background events will correspond to  points at the diagonal 
($\theta_{23} = 180^{\circ} - \theta_{12}$) and below diagonal ($\theta_{23} < 180^{\circ} -\theta_{12}$)
as can be inferred from the middle and left panel of Figure~\ref{fig_scattering}.  
Therefore, one of the possible selection cuts can
be applied on ordered opening angles ($\theta_{12} < \theta_{23} < \theta_{13}$)  between registered gammas, and is
resulting in a decrease of background  by a factor $10^{4}$ while, rejecting only $3\%$  of signal events (see Figure~\ref{fig_theta_cut}).
Combining aforementioned criterion with requirement that registered time difference ($\Delta t$) as a function of detector number ($\Delta ID$) 
is small ($\Delta t < 0.3$~ns), allows for total reduction of the instrumental background
by a factor of $10^{9}$.
However, we have to take into account that the remaining background 
is caused not only by misidentified $2\gamma$ events, but also by 
true annihilations into $3\gamma$ which may originate from the interaction of the 
positronium with surrounding electrons and hence will constitute a background for studies of discrete symmetries. 
Interaction of ortho-positronium with matter is classified into:
pick-off annihilations and ortho-para spin conversion. 
Contribution from these processes depends on the used target material, e.g. 
in aerogel IC3100 and amberlite porous polymer XAD-4 about 7\% and 36\% of ortho-positronium undergo through it, respectively~\cite{Jasinska:2016qsf}.
The events originating from the true of \ops annihilation process ($N_{o-Ps}$) 
can be misidentified with the events from the following processes:
pick-off process with direct annihilation to $3\gamma$
($N_{3\gamma \ pick-off}$);
pick-off process with annihilation to $2\gamma$ misidentified as $3\gamma$ due to
secondary scatterings ($N_{2 \gamma \ pick-off}$); 
conversion of ortho-positronium  to para-positronium with subsequent C symmetry violating decay to $3\gamma$ ($N_{3\gamma \ conv}$);
conversion of ortho-positronium to para-positronium with subsequent annihilation to $2\gamma$ misidentified as $3\gamma$ due to the secondary scatterings ($N_{2\gamma \ conv}$). 

The conservative upper limit of these background contributions may be estimated as:
\begin{align}
    \begin{aligned}
    N_{2\gamma \ conv} / N_{o-Ps} <     N_{2\gamma \ pick-off} / N_{o-Ps} < 
     \\
     N_{3\gamma \ conv} / N_{o-Ps} < N_{3\gamma \ pick-off}  / N_{o-Ps},
    \end{aligned}
\end{align}
where:\\
\begin{minipage}{0.9\columnwidth}
\begin{itemize}
    \item $N_{3\gamma \ pick-off}  / N_{o-Ps}  < (1-\frac{\tau_{matter}}{\tau_{vacuum}}) / 370 \approx 2\cdot 10^{-4} 
        (\mbox{IC3100}) < 10^{-3} (\mbox{XAD-4})$;
    \item $N_{2\gamma \ pick-off} / N_{o-Ps}  <  0.07 \cdot 10^{-9} (\mbox{IC3100}) < 0.36\cdot 10^{-9} (\mbox{XAD-4})$;
    \item $N_{3\gamma \ conv} / N_{o-Ps} < 0.07 \times 2.8 \cdot 10^{-6} (\mbox{IC3100})
        <0.36 \times 2.8 \cdot 10^{-6} (\mbox{XAD-4})$  ;
    \item $N_{2\gamma \ conv} / N_{o-Ps} <  0.07 \cdot  10^{-9}  (\mbox{IC3100}) < 0.36\cdot 10^{-9} (\mbox{XAD-4})$.
\end{itemize}
\end{minipage}
\\
In the above estimations the factor $10^{-9}$ denotes the reduction power of the $2\gamma$ events and 
$2.8\cdot 10^{-6}$ stands for the upper limit of the C symmetry violation via the $\mbox{p-Ps}\rightarrow 3\gamma$
process~\cite{br_measurements}.
The precise control of these contributions will be provided by the measurement of the 
true $2\gamma$ events with high statistics.
\section{J-PET performance in \ops decay measurements}
\label{sec:results}
In order to determine the angular and energy resolution we have performed simulations of ``point-like''
$^{22}$Na source surrounded by water and localized in the geometrical center of the J-PET detector.
The conducted simulations accounted  for positron emission and thermalisation in the target material,
angular and energy distributions of gamma quanta originating from ortho-positronium annihilation 
and Compton interactions of emitted gamma quanta in the J-PET detector. 
Details were presented in the Section~\ref{sec:MonteCarlo}.
In the next step, based on the simulated data, we reconstructed  hit-time and hit-position 
of the registered gamma quantum interaction 
in the detector, taking into account the experimentally determined resolutions. Based on obtained informations
the reconstruction of angles between gamma quanta and of their energies is performed, 
as described in the next paragraph.
\subsection{Angular and energy resolution}
\label{sec:calc_angles}
Incident gamma quantum transmits  energy  as well as momentum to an electron in the plastic scintillator via Compton effect.
Due to that, registered signals at the end of the scintillator strips 
cannot give information about the energy of the incident gamma quantum
on the event-by-event basis.
However, registration of three gamma quanta  hit-position from \ops annihilation
allows reconstruction of their energies based on the energy and momentum conservation.

In CM frame, energies of three gamma quanta from an ortho-positronium annihilation,
can be expressed as a functions of angles ($\theta_{12}, \theta_{23}, \theta_{13}$) between momentum vectors 
(see also Figure~\ref{fig_dalitz}, right panel), as~follows:
\begin{align}
    \begin{aligned}
        E_1 &= - 2m_e\frac{- \cos \theta_{13} +  \cos\theta_{12} \cos\theta_{23}}{(-1 + \cos\theta_{12}) (1 + \cos\theta_{12} - \cos\theta_{13} - \cos\theta_{23})}, \\
        E_2 &= - 2m_e \frac{ \cos\theta_{12} \cos\theta_{13} - \cos\theta_{23}}{(-1 + \cos\theta_{12}) (1 + \cos\theta_{12} - \cos\theta_{13} - \cos\theta_{23})}, \\
        E_3 &= 2 m_e \frac{1 + \cos\theta_{12}}{1 + \cos\theta_{12} - \cos\theta_{13} - \cos\theta_{23}}.
        \label{eq_energy}
    \end{aligned}
\end{align}
The measured positions of gamma interaction in the detector, together with known or reconstructed position 
of annihilation, allow for $E_i$ determination.
The determination of angles requires reconstruction of interaction points and annihilation position. 
As regards annihilation position we may distinguish two cases, discussed in the next paragraphs.
\subsubsection{Point-like positronium source}
\label{sec:resolution}
In some cases of discrete symmetries studies positronium  will be produced in the well localized material
surrounding the ``point-like'' positron source~\cite{Moskal:2016moj}. 
Assuming that $\beta^+$ emitter position corresponds to the ortho-positronium annihilation point,
the angles ($\theta_{12}$, $\theta_{13}$ and $\theta_{23}$) between gamma quanta can be determined
from registered gamma quanta interaction points ($\vec{r}_{hit}$) in the detector. 
Coordinates $x$ and $y$ are determined as the centre of the scintillator strip,
and therefore the precision of their determination correspond  to the geometrical cross section 
of the scintillator strip. The $z$ coordinate is determined from 
signals arrival time to photomultipliers at the ends of scintillator strip, and its uncertainty 
is equal to about $\sigma(z) = 0.94$~cm~\cite{Raczynski:2015zca,Raczynski:2014poa}. 
Uncertainty of $\sigma(\vec{r}_{hit})$ determination gives the main contribution to 
estimation of angular and energy resolutions. The second order effect is an
uncertainty originating from non zero boost and distance traveled by positron in matter.
\begin{figure*}
    \centering
    \begin{subfigure}[b]{\textwidth}
        \centering
        \includegraphics[width=0.48\textwidth]{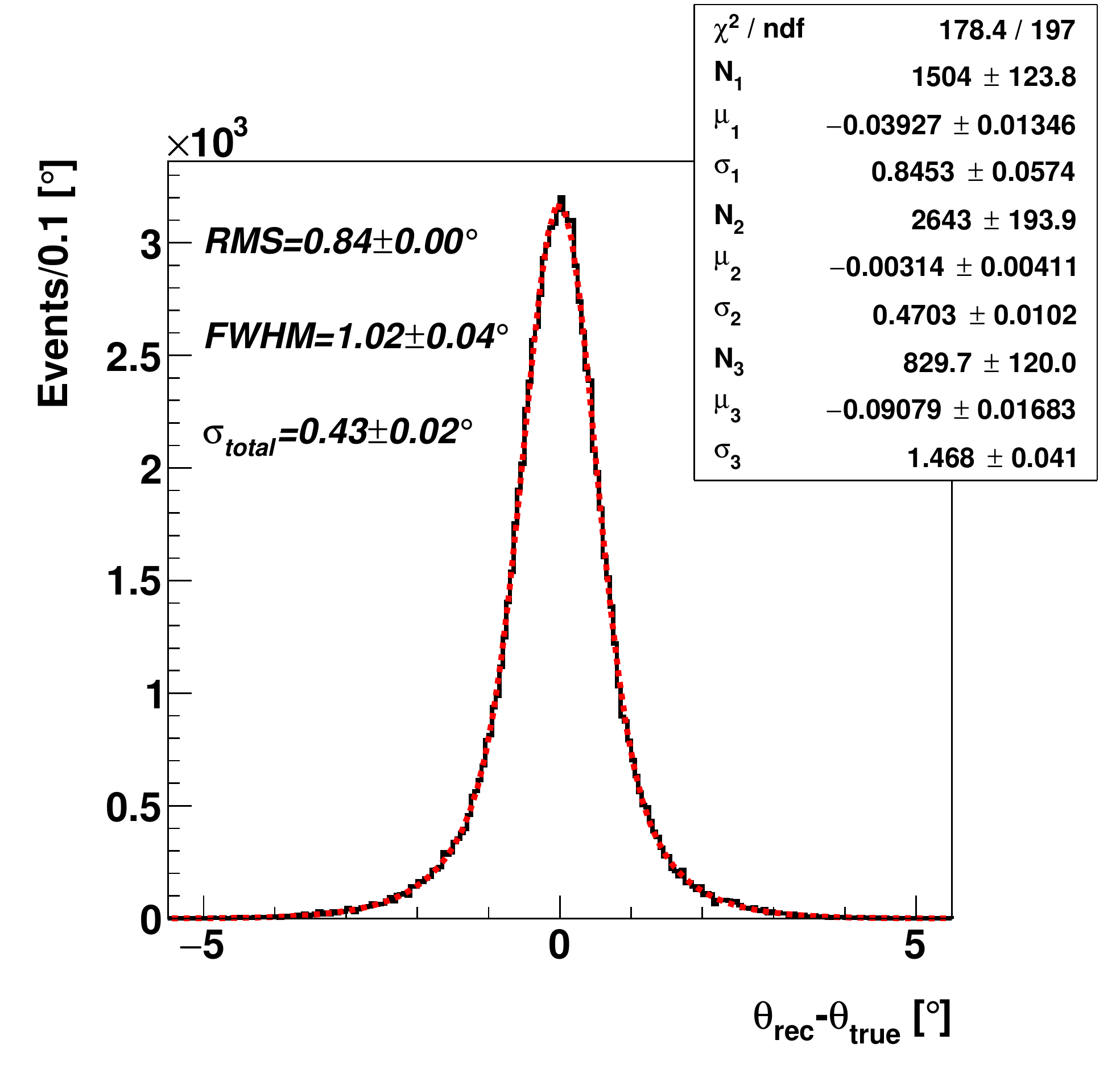}
        \includegraphics[width=0.48\textwidth]{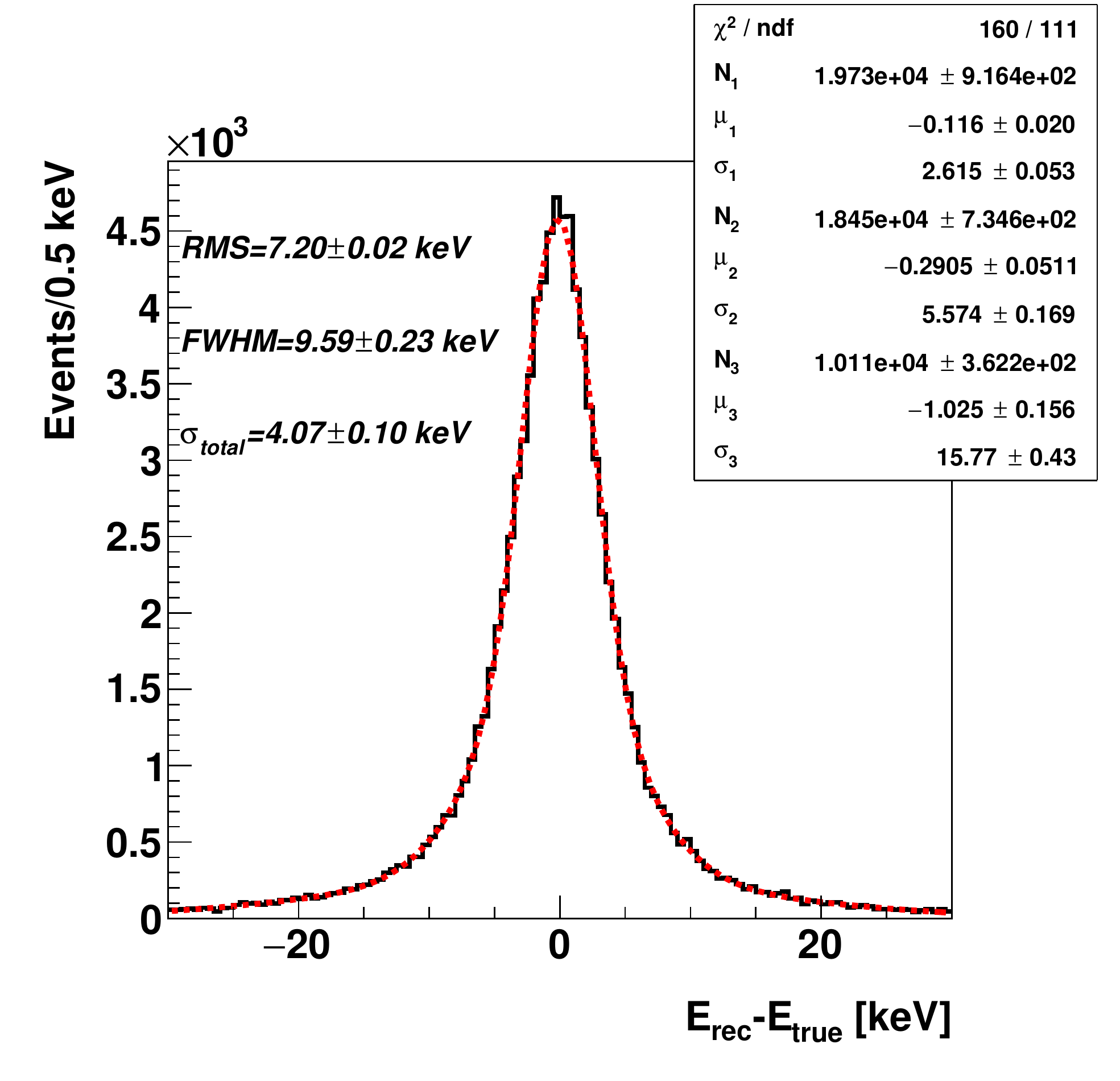}
    \end{subfigure}
    \caption{Resulting angular (left) and energy (right) resolution 
        spectra for ``point-like'' positronium source with known location 
    and assumed detector resolution $\sigma(T^0_{hit}) = 80$~ps.}
    \label{fig_resolution_fit}
\end{figure*}
\begin{figure*}
    \centering
    \begin{subfigure}[b]{\textwidth}
        \centering
        \includegraphics[width=0.48\textwidth]{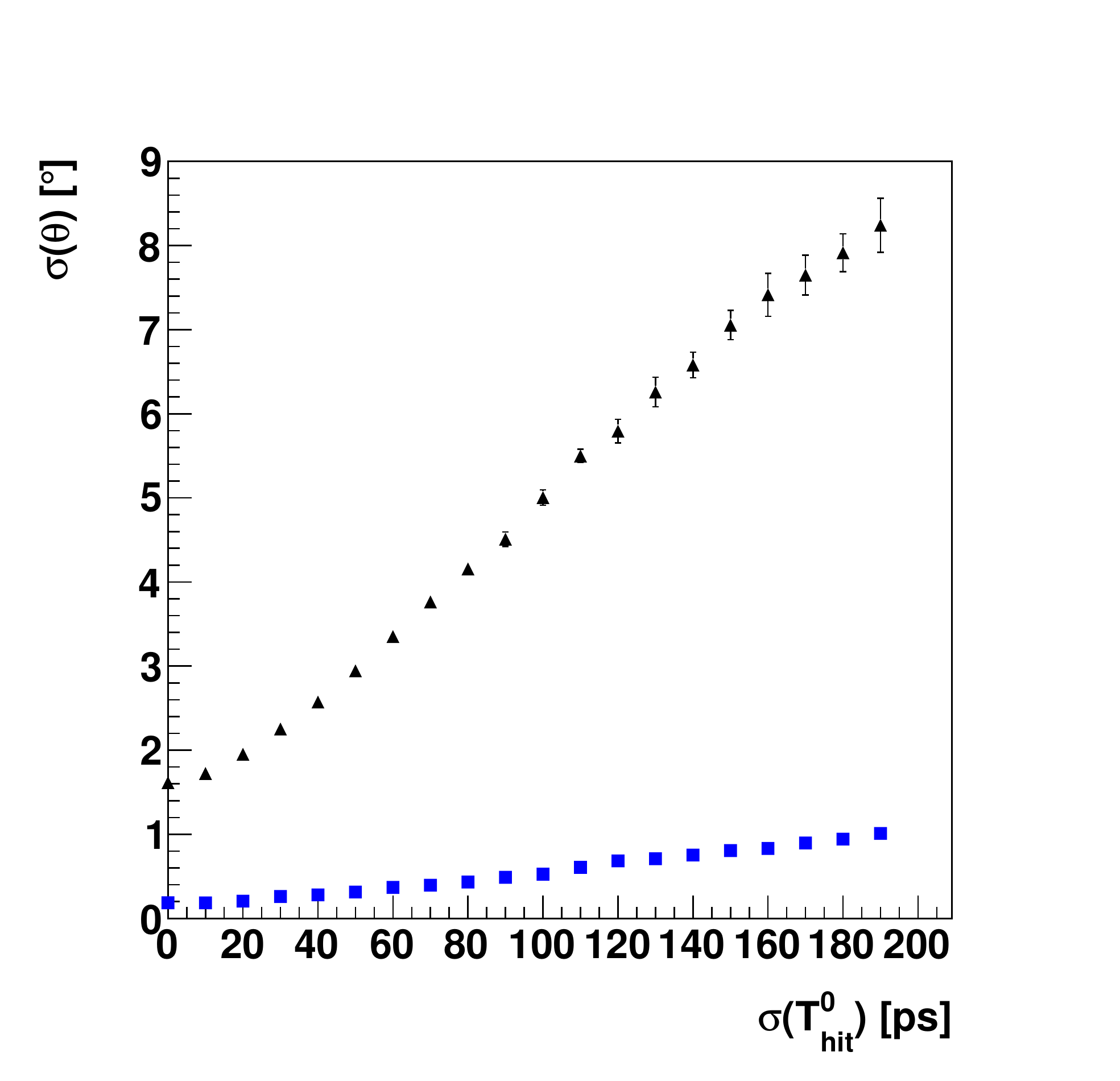}
        \includegraphics[width=0.48\textwidth]{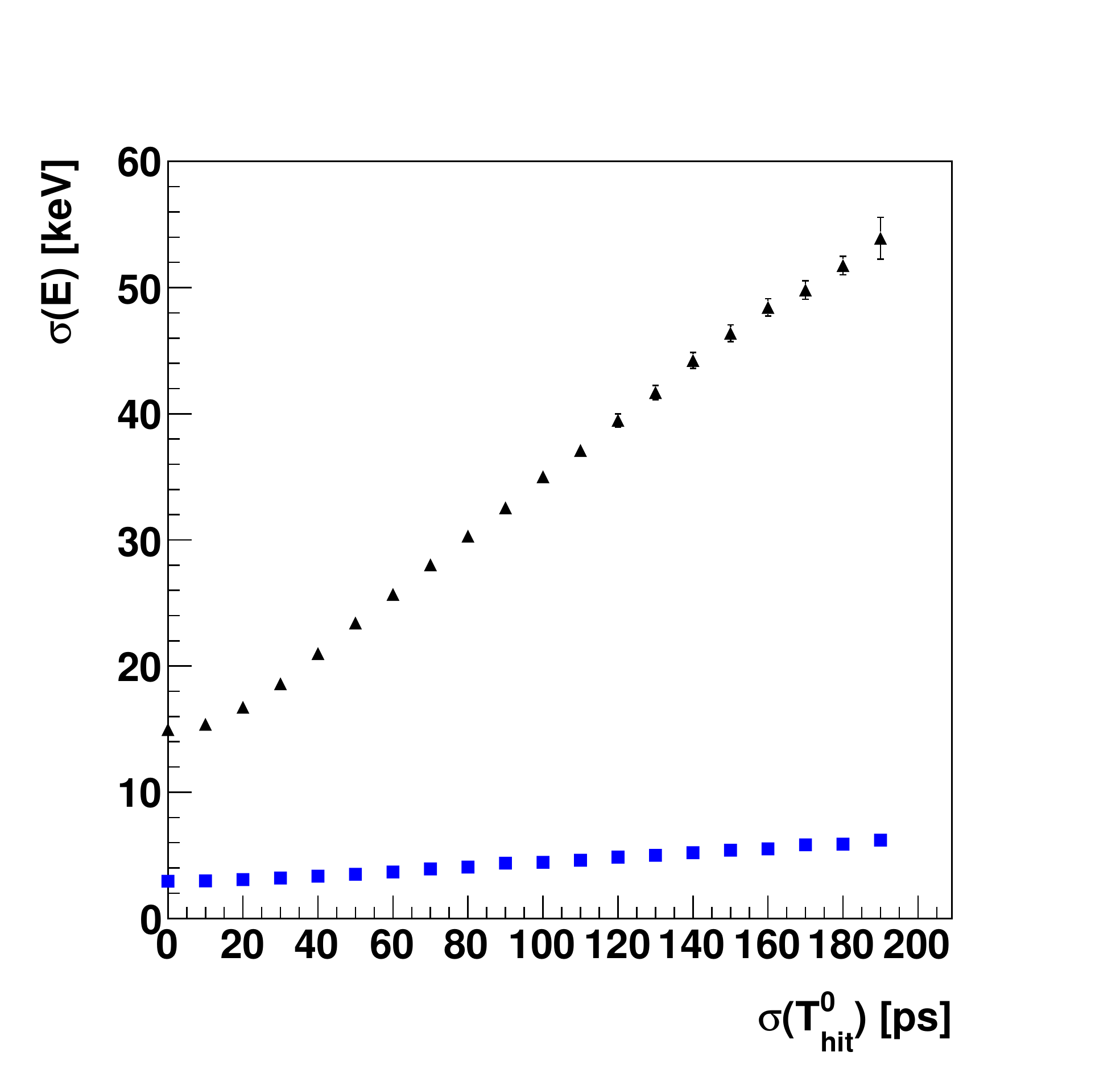}
    \end{subfigure}
    \caption{
        Angular (left) and energy (right) resolution for the registration of the gamma quanta originating from ortho-positronium annihilation 
        as a function of detector time resolution for ``point-like'' (blue box) and extended (black triangle) 
    positronium source.}      
    \label{fig_resolution}
\end{figure*}

\subsubsection{Spatially extended positronium source}
\label{sec:rec_algo}
The angles ($\theta_{12}$, $\theta_{23}$, $\theta_{13}$) and hence a full kinematics of \ops decay 
can be also reconstructed in the case of the extended positronium target.
For example a target of a cylindrical shape with the diameter of 20~cm  
was proposed  for the production 
of a linearly polarized  positronium~\cite{Moskal:2016moj}.
Polarisation can be determined provided that positron emission and positronium formation 
(approximately the same as annihilation) position are known.

A new reconstruction algorithm that allows  reconstruction 
of ortho-positronium annihilation position for an event by event basis
was recently reported~\cite{Gajos:2016nfg,patent}. 
The method  based on trilateration allows for a simultaneous reconstruction of both location and time of the
annihilation based on  time and  interaction position of gamma quanta in the J-PET detector.
The reconstruction performance strongly depends on detector time resolution ($\sigma(T_{hit})$).
Using aforementioned reconstruction algorithm, 
current J-PET spatial resolution for annihilation  point reconstruction is at the
level of 1.5~cm along the main detector axis and 2~cm in the transverse plane~\cite{Gajos:2016nfg}.
\subsubsection{Performance studies}
The angular and energy resolutions for the registration of the gamma quanta from the \ops decay
are established from simulations i.e. the distributions of the differences between generated and 
reconstructed values of angles and energies.
Figure~\ref{fig_resolution_fit}  show results obtained under assumption that the hit-time 
resolution is given by equations~\ref{sigma_E} and \ref{sigma_E_lower_energy}.
In order to determine the angular and energy resolution the triple Gaussian model,
which effectively describes obtained distributions, was applied:
\begin{equation}
    f(x) = \sum_{i=1}^{3} \frac{N_i}{\sqrt{2\pi} \sigma_i} \cdot e^{-\frac{1}{2} \left(\frac{x-\mu_i}{\sigma_i}\right)^2}, 
\end{equation}
where $N_i$, $\mu_i$ and $\sigma_i$ were varied in the fit.
The total uncertainty was obtained as a standard deviation of the total distribution equivalent~to:
\begin{equation}
    \sigma_{total} = \sqrt{ \sum_{i=1}^{3} \left[ \left(\frac{N_i}{\sum_{j=1}^{3} N_j}\right) \cdot \sigma_{i}  \right]^2}.
    \label{eq_sigma_total}
\end{equation}

Since the angular and energy resolution strongly depend on hit-time resolution registered 
in the J-PET detector, the studies of resolution  were made for $\sigma(T_{hit}^0)$ in 
the range
from 0~ps to 190~ps.
Comparison between obtained resolutions for the ``point-like'' and extended positronium source 
is shown in Figure~\ref{fig_resolution}.
In both cases energy and angular resolutions are improving with decreasing 
$\sigma(T_{hit}^0)$, and for presently achieved time resolution of $\sigma(T^0_{hit})$,
and well a localized ``point-like'' positronium source, they amount to $\sigma(\theta) = \sigmaThetaPointLike$ and
$\sigma(E_{hit}) = \sigmaEnergyPointLike$, respectively.
In case of the extended positronium source, when the reconstruction  of the annihilation point is needed both 
resolutions increases to $\sigma(\theta) = \sigmaThetaGps$ 
and $\sigma(E_{hit}) = \sigmaEnergyGps$, respectively.
\subsection{J-PET efficiency studies with Monte Carlo simulations}
The rate of registered \ops  events in general can be expressed by the formula:
\begin{equation}
    R_{oPs\rightarrow 3\gamma} = A \cdot f_{oPs\to 3\gamma} \cdot  \epsilon_{det}(th) \cdot \epsilon_{ana}, 
\end{equation}
where $A$ is the total annihilation rate 
(fast timing of applied  plastic scintillators allows for usage of the 10~MBq positron source), 
$f_{oPs\to 3\gamma}$ is the fraction of annihilations via \ops process
in the target material, $\epsilon_{det}(th)$
is the detector efficiency as a function of applied detection threshold
while $\epsilon_{ana}$ denotes selection efficiency used
to discriminate between $3\gamma$ and $2\gamma$ events.

\begin{figure*}[t]
    \centering
    \begin{subfigure}[b]{0.40\textwidth}
        \centering
        \includegraphics[width=\textwidth]{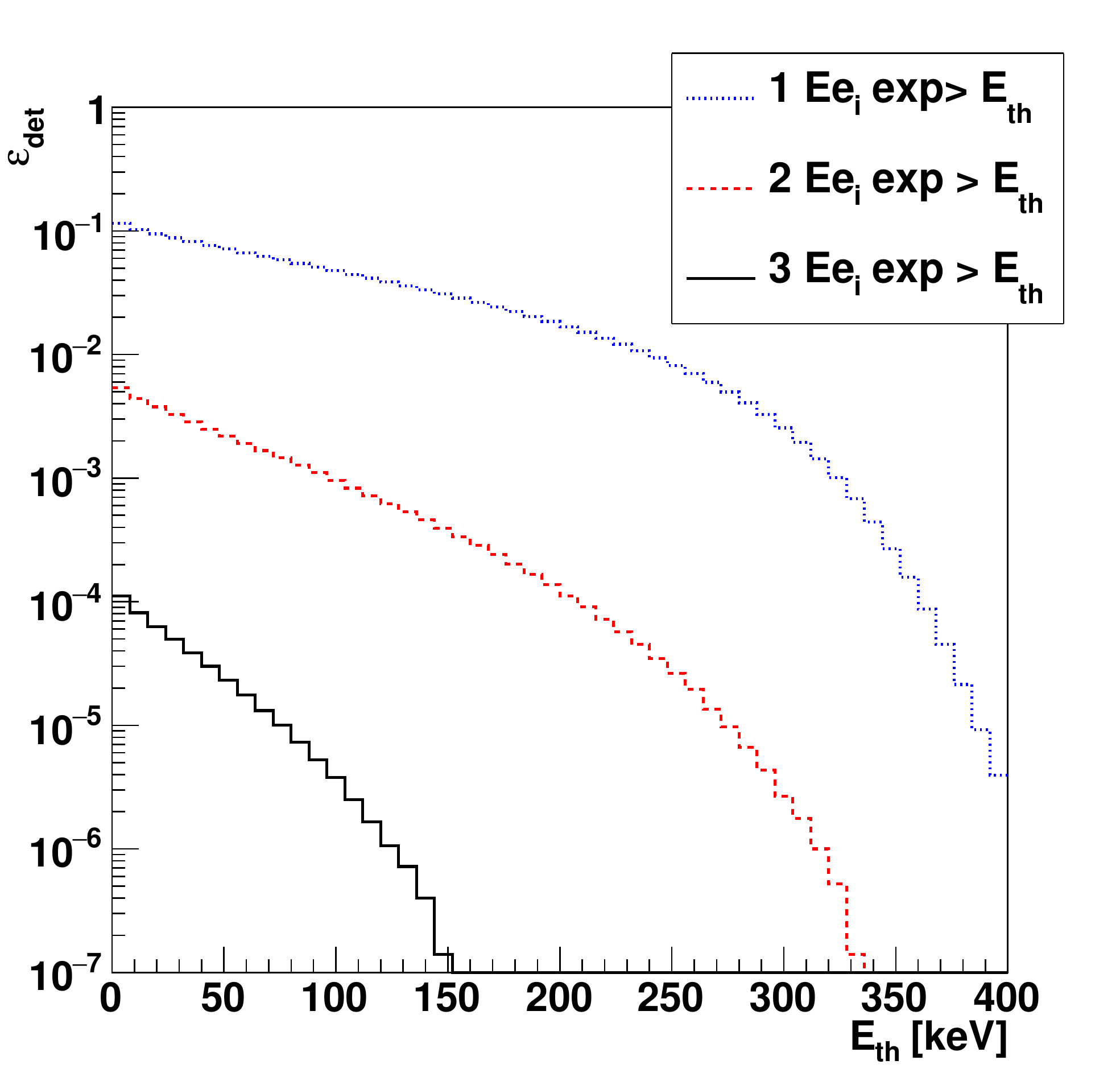}
        \caption{ J-PET  }
        \label{fig_effi_real_detector}
    \end{subfigure}
    \begin{subfigure}[b]{0.40\textwidth}
        \centering
        \includegraphics[width=\textwidth]{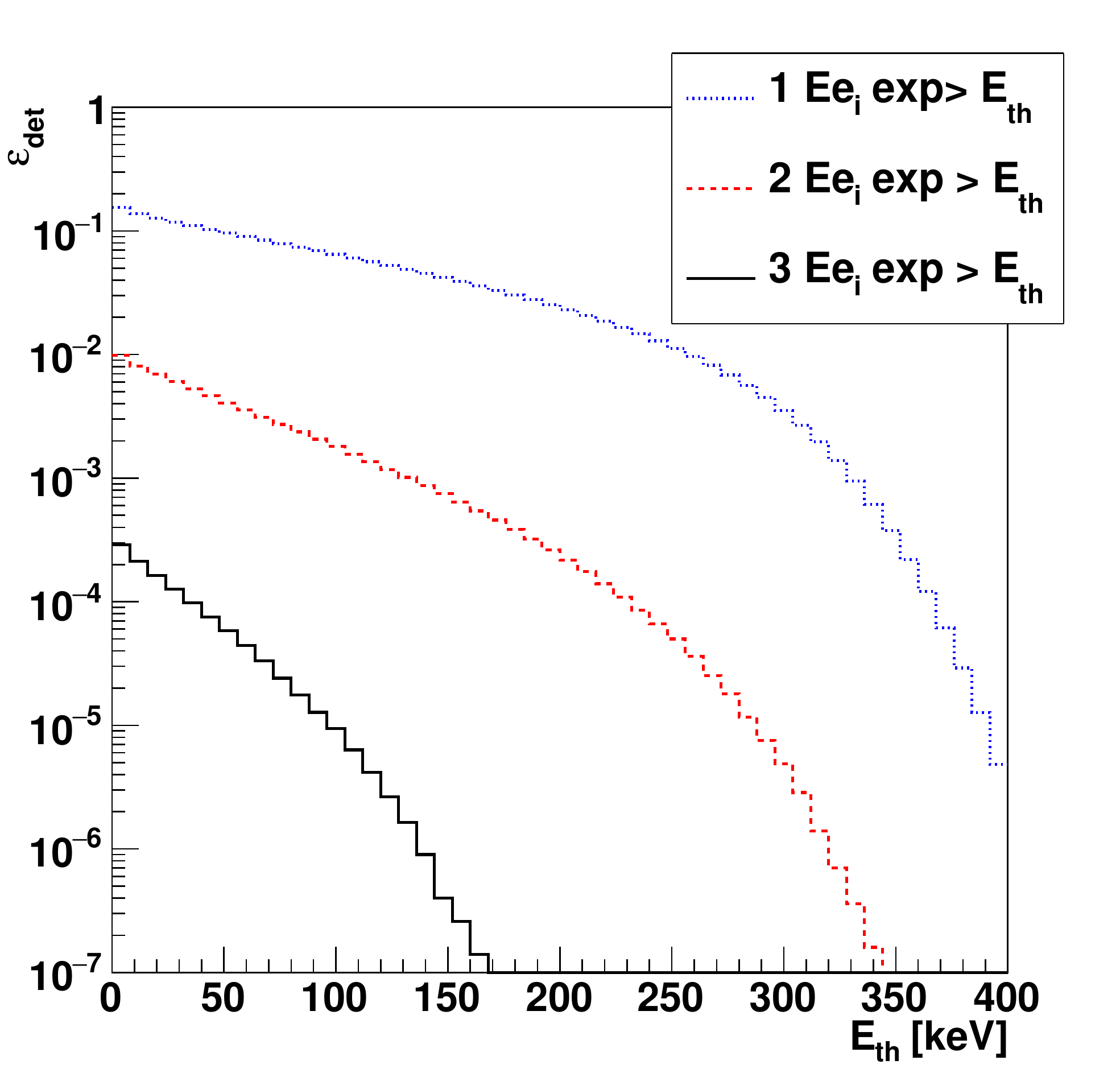}
        \caption{ J-PET+1}
        \label{fig_effi_real_detector_with_extra_layer}
    \end{subfigure}
    \begin{subfigure}[b]{0.40\textwidth}
        \centering
        \includegraphics[width=\textwidth]{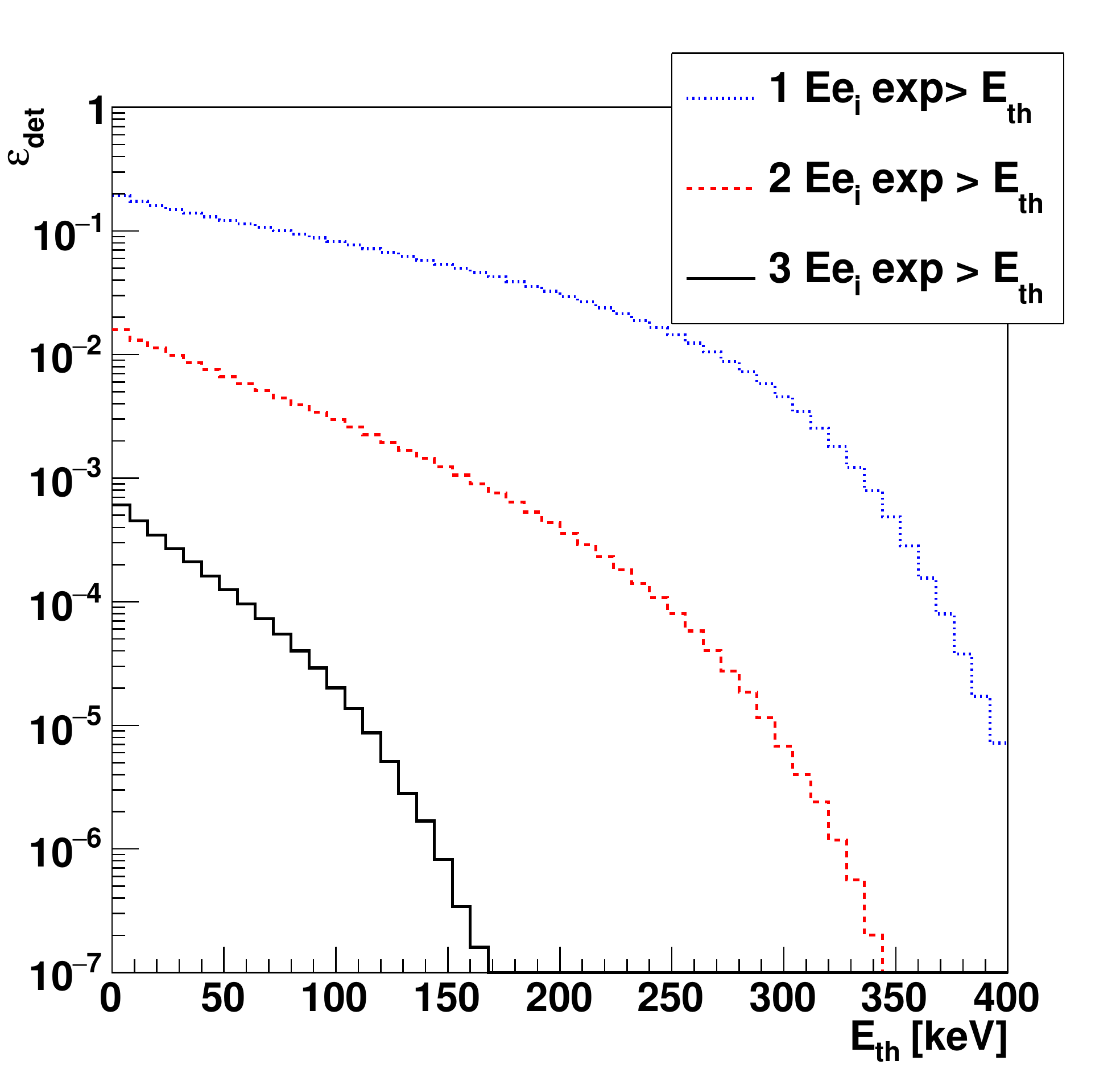}
        \caption{ J-PET+2  }
        \label{fig_effi_real_detector_with_2extra_layer}
    \end{subfigure}
    \begin{subfigure}[b]{0.40\textwidth}
        \centering
        \includegraphics[width=\textwidth]{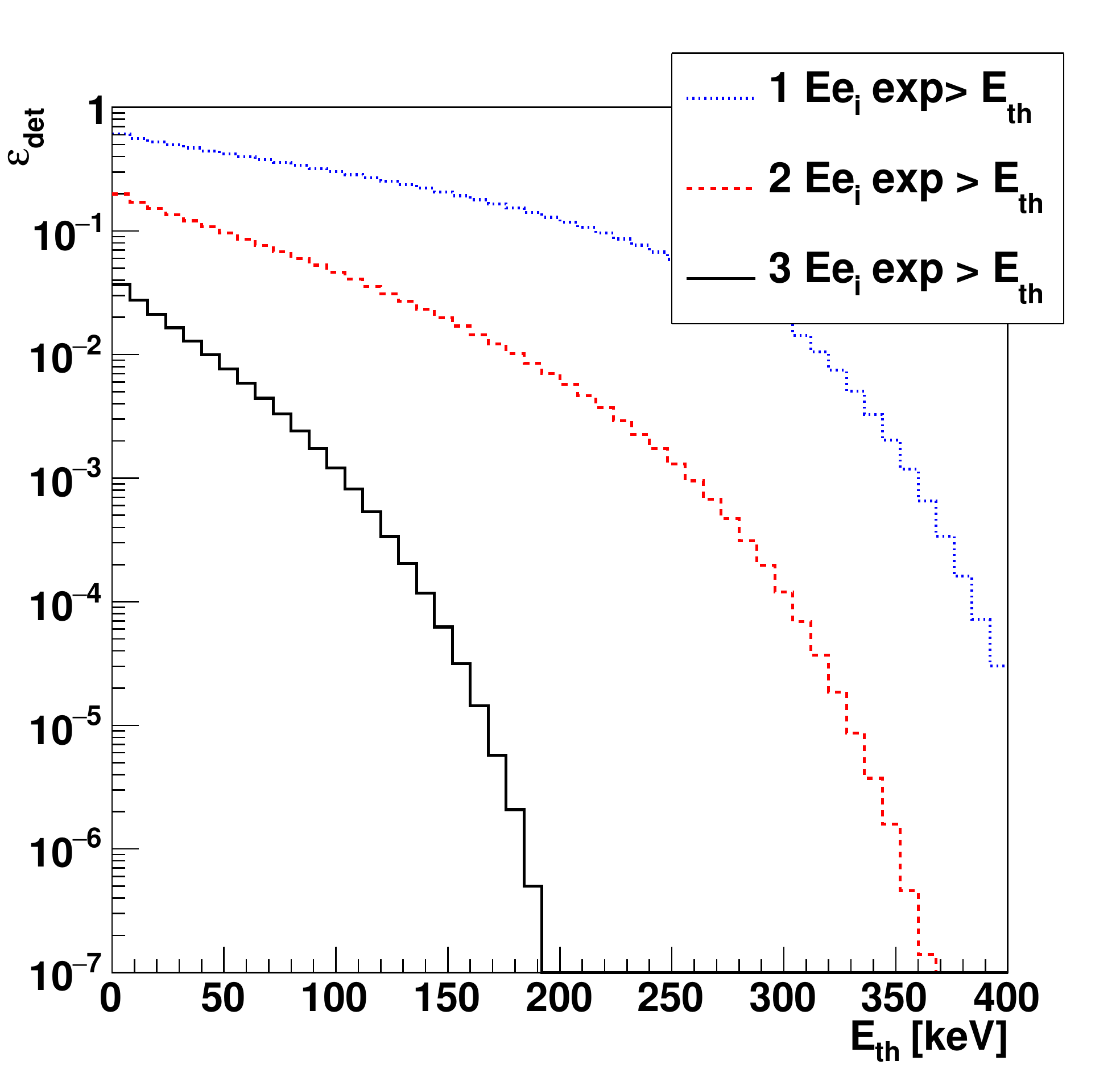}
        \caption{ J-PET-full  }
        \label{fig_effi_ideal_detector}
    \end{subfigure}
    \caption{$\mbox{oPs} \rightarrow 3 \gamma$ registration efficiency (determined
        taking into account 
        geometrical acceptance, 
        probability of gamma quanta registration in the plastic scintillator 
        and J-PET detector resolution)
        as a function of applied threshold
        for different types of simulated geometries.
        The shown dotted, dashed and solid lines  indicate efficiency assuming that at least one, two  or three  photons 
        deposited energy above the threshold, respectively.  
    }
    \label{fig_effi}
\end{figure*}

The $\epsilon_{det}$ efficiency of the \ops reconstruction will depend on the
energy deposition threshold used
in the analysis (see Figure~\ref{fig_exp_ve_theory}).
The hardware threshold at the order of 10~keV~\cite{Moskal:2014sra}
will be set to discriminate the experimental noise and later on we will apply further selection threshold based on the measured energy deposition.
The probability of registration of 1, 2 or 3 gamma quanta originating from \ops annihilation ($\epsilon_{det}$) 
as a function of applied selection threshold in different geometries is shown in Figure~\ref{fig_effi}. 
Efficiency $\epsilon_{det}$ contains contribution from geometrical acceptance, probabilities of gamma quanta
interaction in applied plastic scintillators and it was determined taking into account the J-PET detector resolution. 
In our evaluation we assume conservatively that the event selection threshold will be set to 50~keV.
A fraction of annihilations via \ops process is estimated taking into account only
longest lived component in two selected materials
IC3100 ($f_{oPs \to 3\gamma} = 16.6\%$) and XAD-4 ($f_{oPs \to 3\gamma} =28.6\%$)~\cite{Jasinska:2016qsf}.
The expected rate of registered signal events is shown in Table~\ref{tab_ratio}.
Using in the experiment amberlite porous polymer XAD-4 instead of aerogel IC3100 as target material, allows to collect
the required statistics almost twice faster,
however, resulting  with higher systematic uncertainties due to the interaction of positronium with the target material, 
as discussed in Section~\ref{sec:bck_rejection}. 
\begin{table}
        \caption{Expected rate of registered signal events in different geometries and target materials 
            assuming $10^6$ annihilations per second and requiring energy deposition above 50~keV for all three gamma quanta 
        from \ops decay.}
        \label{tab_ratio}
        \begin{tabular*}{\columnwidth}{@{\extracolsep{\fill}}|c||cccc|@{}} \hline
            \multirow{2}{*}{Target material} & \multicolumn{4}{c|}{Rate  of registered \ops events  $\left[s^{-1}\right]$ } \\ \cline{2-5} 
                                             & J-PET & J-PET+1 & J-PET+2 & J-PET-full \\ \hline 
            IC3100 & 15 & 70 & 130 & 10600 \\
            XAD-4 & 25 & 115 & 230 & 18300  \\ \hline
        \end{tabular*}  
\end{table}
 
\section{Conclusions}
\label{sec:conclusions}
We presented results of Monte Carlo simulations showing that the Jagiellonian-PET multipurpose detector constructed at the Jagiellonian 
University allows exclusive registration of the decays of ortho-positronium  into three photons (o-Ps$\rightarrow 3 \gamma$)
providing angular and energy resolution of $\sigma(\theta) \approx \sigmaThetaPointLike$ and $\sigma(E) \approx \sigmaEnergyPointLike$, respectively.

The achieved results indicate that the J-PET detector gives a realistic chance to improve the best
present limits established for the  CP and CPT symmetry  violations 
in the decays of positronium~\cite{gammasphere,tokyo} by more than an order of magnitude.
This can be achieved by (i) collecting 
at least two orders of magnitude higher statistics, due to the possibility of
using a $\beta^+$ source with higher rate (10~MBq at J-PET vs 0.37~MBq at
Gammasphere~\cite{gammasphere} or 1~MBq
at Tokyo University experiment~\cite{tokyo}), 
(ii) the enhanced fraction of $3\gamma$ events by the use of the amberlite polymer XAD-4,
(iii) a measurements with a few times improved angular resolution and
(iv) about two times higher degree of o-Ps polarization, as shown recently in reference~\cite{Gajos:2016nfg}.
The limitation on the source activity can be overcome by the J-PET due to the application of 
plastic scintillators that are characterized by about two orders 
of magnitude shorter duration of signals, 
thus decreasing significantly the pile-ups problems with respect to the crystal based detector systems.
In addition, the improved angular resolution combined with the superior timing of the J-PET detector
(by more than order of magnitude improved with respect to the crystal detectors)
and with the possibility of the triggerless registrations~\cite{Korcyl:2014,Korcyl:2016pmt} 
of all kind of events with no hardware coincidence window 
allow  suppression and monitoring of the background, due to 
misidentification of $2\gamma$ events and  possible contribution from $3\gamma$ pick-off annihilations.

\section*{Acknowledgments}
We acknowledge valuable discussions with Dr J.~Wa\-wry\-szczuk and technical
and administrative support by A.~He\-czko, M.~Kajetanowicz, W.~Migdał,
and the financial support by the Polish National Center for Research and
Development through grants INNOTECH-K1/IN1/64/159174/NCBR/12
and LIDER\--274/L-6/14/NCBR/2015, the Foundation for Polish Science
through MPD program and the EU, MSHE Grant No. POIG .02.03.00-161
00-013/09, Marian Smoluchowski Kraków Research Consortium ``Matter–Energy–Future'',
and the Polish Ministry of Science and Higher Education
through grant 7150/E-338/M/2015. B.C.H. gratefully acknowledges the
Austrian Science Fund FWF-23627.

\end{document}